\documentclass[reprint,aps,amssymb,amsmath]{revtex4-1}
\pdfoutput=1 
\usepackage{graphics} 
\usepackage{graphicx} 

\usepackage{color}

\begin{document} 
\renewcommand{\i}{\operatorname{i}}

%\title{Computing Gibbs free energy differences between phases \\from the thermodynamic force needed to pin interfaces} 
%\title{Computing the crystal growth rate by interface pinning:\\a method where almost all constants are name $\mu$}
\title{Computing the crystal growth rate by the interface pinning method}
%\title{Ab initio computations of the crystal growth rates using interface pinning}
%\email{ulf.pedersen@tuwien.ac.at} 

\author{Ulf R. Pedersen$^{1,2}$} 
\email{ulf@urp.dk} 
\author{Felix Hummel$^{3}$} 
%\author{Georg Kresse$^{2}$} 
%\author{Gerhard Kahl$^{1}$} 
\author{Christoph Dellago$^{3}$} 
\affiliation{$^1$Department of Sciences, Roskilde University, P. O. Box 260, DK-4000 Roskilde, Denmark} 
\affiliation{$^2$Institute of Theoretical Physics, Vienna University of Technology, Wiedner Hauptstra{\ss}e 8-10, A-1040 Vienna, Austria}
\affiliation{$^3$Faculty of Physics, University of Vienna and Center for Computational Materials Science, Sensengasse 8/12, A-1090 Vienna, Austria} 

\date{\today} 
%\pacs{...} 
%\keywords{molecular dynamics simulation, phase diagrams, computing Gibbs free energy}

\begin{abstract} 
An essential parameter for crystal growth is the kinetic coefficient given by the proportionality between super-cooling and average growth velocity. 
Here we show that this coefficient can be computed in a single equilibrium simulation using the interface pinning method where two-phase configurations are stabilized by adding an spring-like bias field coupling to an order-parameter that discriminates between the two phases. Crystal growth is a Smoluchowski process and the crystal growth rate can therefore be computed from the terminal exponential relaxation of the order parameter. The approach is investigated in detail for the Lennard-Jones model. We find that the kinetic coefficient scales as the inverse square-root of temperature along the high temperature part of the melting line. The practical usability of the method is demonstrated by computing the kinetic coefficient of the elements Na, Mg, Al and Si from first principles. It is briefly discussed how a generalized version of the method is an alternative to forward flux sampling methods for computing rates along trajectories of rare events.
\end{abstract}

\maketitle 

\section{Introduction}

Crystal growth is of paramount importance in many branches of condensed matter physics \cite{pimpinelli1998,woodruff1973}. 
A important parameter in phase field equations \cite{provatas2010} describing crystal growth is the kinetic coefficient defined as the proportionality constant between super cooling (or super heating) and the average interface growth (or melting) velocity $\langle \dot{x}_s\rangle$ \cite{hoyt2002kc,hoyt2002}. We define the kinetic coefficient $M$ using the difference in chemical potential between the two phases $\mu_{sl}=\mu_s-\mu_l$ as a measure of super cooling (or super heating):
\begin{equation} \label{kc} 
 \langle \dot{x}_s \rangle = -M\mu_{sl}
\end{equation} 
In the spirit of the fluctuation-dissipation theorem \cite{kubo1966} we suggests to learn about the interface dynamics by investigating spontaneous fluctuations when a bias potential is added to the Hamiltonian.
Specifically, we determine $M$ indirectly in a computation where trajectories are pinned to two-phase configurations (Fig. \ref{ip_scetch}) by adding a harmonic energy coupling to an order-parameter of crystallinity \cite{physrevB2013,pedersen_ip2013}. In effect, non-equilibrium crystal growth is converted into a well-defined equilibrium problem. We devise a stochastic model of fluctuations that assumes Smoluchowski dynamics for crystal growth and show that the growth rate can be inferred from the terminal exponential relaxation of the order-parameter. The chemical potential difference between the solid and the liquid $\mu_{sl}$ is known from the average force exerted by the bias field on the system, and thus the proportionality constant $M$ can be computed in a single equilibrium simulation.

The paper is organized as follows: Sec. \ref{secIP} is a brief introduction to the interface pinning method. In Sec. \ref{secSM} we motivate and give the solution to a stochastic model of thermal fluctuations with and without a pinning potential. In Secs. \ref{secLJ} and \ref{secDFT} we compute crystal growth rates for the Lennard-Jones model and combine the method with {\it ab initio} DFT to compute growth rates of the elements Na, Mg Al and Si. The paper is finalized with a discussion of the method (Sec. \ref{secDiscussion}). The Appendix gives a detailed analysis of our stochastic model.

%In an elongated periodic box \cite{physrevB2013,pedersen_ip2013} we define the interface position $x_s$ as shown on Fig. \ref{ip_scetch}.

\section{Interface pinning}\label{secIP}

\begin{figure} 
\begin{center} 
  \includegraphics[width=0.5\columnwidth]{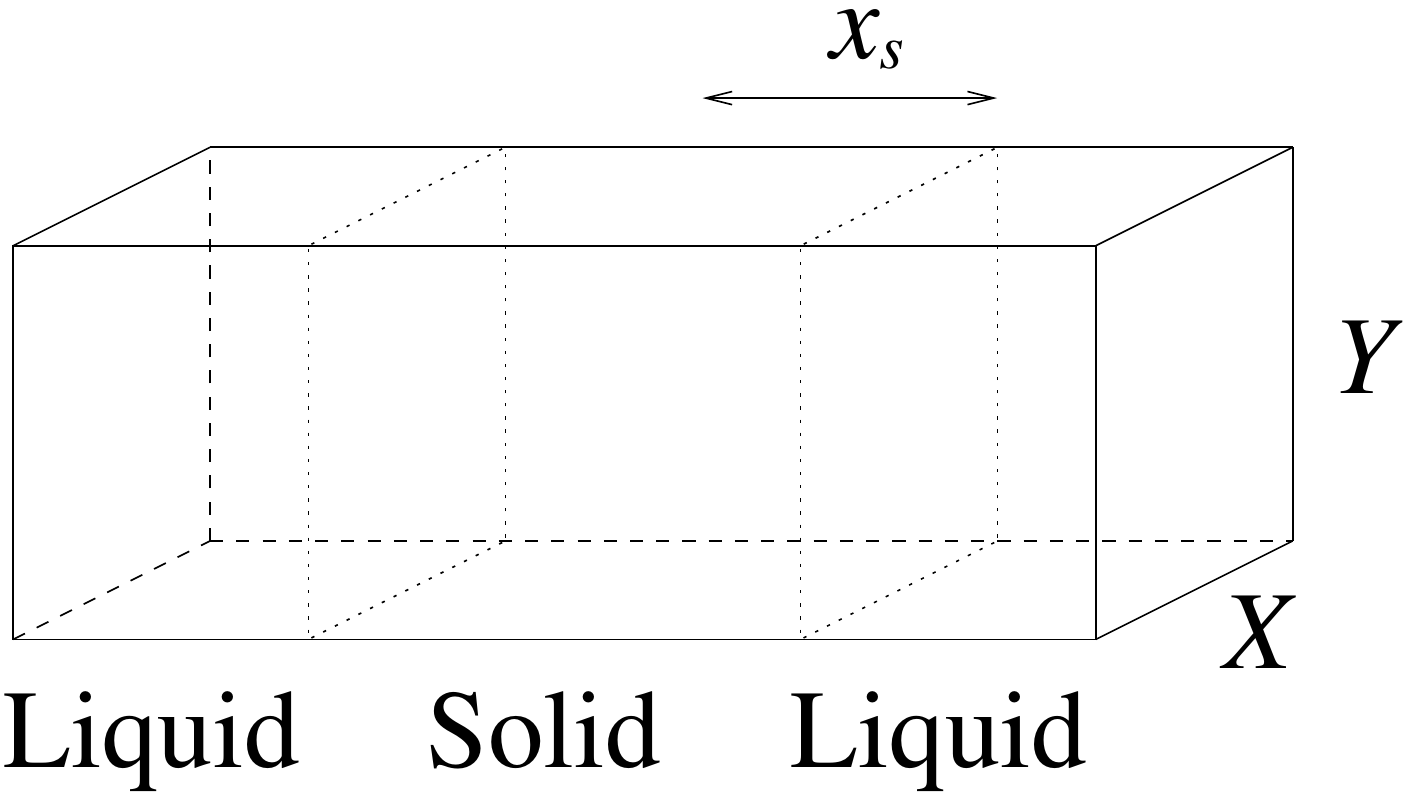} 
  \caption{\label{ip_scetch} Sketch of an elongated periodic simulation box containing a two-phase configuration. The projected interface position, ignoring capillary waves, is $x_s=N_s/2XY\rho_s$.} 
\end{center} 
\end{figure}

In the following we give a brief review of the recently proposed interface pinning method \cite{physrevB2013,pedersen_ip2013} for studying the solid-liquid phase transition. Consider a system of $N$ particles in an elongated periodic simulation cell as sketched in Fig. \ref{ip_scetch}. 
For the configuration $\mathbf{R}=(\mathbf{r}_1,\mathbf{r}_2,\ldots,\mathbf{r}_N)$ let the system have the Hamiltonian ${\mathcal H}_0(\mathbf{R})$. Hydrostatic pressure is ensured by allowing the box length in the $z$-direction to change and constant temperature is ensured by connecting the momenta of the particles to a heat bath. We refer to this as the $Np_zT$-ensemble. When the initial $\mathbf{R}$ is a two-phase configuration, the thermodynamically stable phase will grow on the expense of the other phase. In an interface pinning computation this is avoided by adding an auxiliary spring-like energy term with a spring constant $\kappa$ and an anchor point $\bar{Q}$ to the Hamiltonian so that $\mathbf{R}$ is biased towards two-phase configurations:
\begin{equation}
\label{eq:hamilton}
{\mathcal H}(\mathbf{R})= {\mathcal H}_0(\mathbf{R}) + \frac{\kappa}{2}[Q(\mathbf{R})-\bar Q]^2.
\end{equation}
Here $Q(\mathbf{R})$ is a measure of crystallinity of the system. In practice we use long-range order as measured by the magnitude of the collective density field $|\rho(\mathbf{k})|$ \cite{pedersen_ip2013} (unless otherwise stated).
\begin{equation}\label{eq:rhok}
Q(\mathbf{R})=|\rho(\mathbf{k})|=\left|\sum_i^N\exp(-\i \mathbf{k}\cdot\mathbf{r}_i)\right|,
\end{equation}
where $\mathbf{k}$ is a wavevector that corresponds to a Bragg peak.
The averaged force $\alpha$ the system exerts on $Q$ is proportional to the chemical potential difference between the two phases:
\begin{equation}\label{eq:alpha}
\alpha = N\mu_{sl}/Q_{sl}.
\end{equation}
Here $Q_{sl}=Q_s-Q_l$, and $Q_s$ and $Q_l$ are the mean values of the order parameter when the system is entirely crystalline or liquid, respectively.
When the system is in equilibrium with respect to $\mathcal H$, the relative position of the interface stops moving up to thermal fluctuations and the force $\alpha$ is balanced by the average force $\kappa(\langle Q\rangle - \bar Q)$ of the spring-like bias-field. Thus, $\mu_{sl}$ can be computed from the average spring force as follows:
\begin{equation}\label{eq:musl}
\mu_{sl} =  Q_{sl}\kappa(\langle Q\rangle - \bar Q)/N.
\end{equation}
% 
%Coexistence ($\mu_{sl}=0$) can be determined using Newtons root finding method. The required derivatives are given by standard thermodynamic relations. 
%
Since we have used a harmonic pinning potential, the distribution of the order parameter will be Gaussian with variance $k_BT/\kappa$. The time evolution of the order parameter $Q(t)$ depends on the trajectory $\mathbf{R}(t)$ that itself is determined by ${\mathcal H}_0$. The main idea of the method suggested in this paper is that if we can understand the time fluctuations of $Q(t)$ when the interface pinning field is applied, we can deduce how the system evolves in the absence of the interface pinning potential. In the following section we devise a stochastic model that assumes that crystal growth is an over-damped Smoluchowski process. From this model we deduce the value of the kinetic growth coefficient $M$.

\section{Stochastic model} \label{secSM}
This section suggests a simple model for the dynamics of $Q(t)$ that enables us to determine the value of $M$ from an interface pinning simulation.
Assuming Newtonian dynamics of $\mathbf{R}$, the $Q(t)$-trajectory is deterministic and uniquely defined from the initial values of $\mathbf{R}$ and $\dot{\mathbf{R}}$. We are, however, only interested in the statistical behavior of $Q(t)$, e.g., the autocorrelation function $\langle \Delta Q(0)\Delta Q(t)\rangle$. To this aim, we devise a stochastic model with an effective Hamiltonian that is a function of coarse-grained coordinates. The equations of motion of these coordinates include Langevin noise forces representing degrees of freedom that have been neglected \cite{langevin1908}. The essential model parameter for determining $M$ is ``the friction constant of crystal growth'' $\gamma$. We determine the value of this parameter by fitting the model to simulation results.

\subsection{Effective Hamiltonian}
As mentioned earlier, the interface pinning method requires the definition of an order parameter $Q({\bf R})$ that quantifies how much of the system is crystalline at the configuration ${\bf R}$, which gives the positions of all atoms. Ideally, the order parameter $Q$ should be proportional to the number $N_{\rm c}$ of crystalline particles such that any increase in $Q$ corresponds to a growth of the crystalline region. In practice, however, the crystallinity of the system is measured using the Fourier transform of the density field as order parameter, Eq. (\ref{eq:rhok}). In this case, $Q$ changes not only due a decrease or increase of the number of crystalline particles, but also due to other spatial fluctuations such as lattice vibrations. Thus, we write the order parameter $Q$ as a sum of two contributions,
\begin{equation}
Q(t)=q(t) + f(t),
\end{equation}
where $q$ is assumed to be proportional to the number of crystalline atoms and $f$ takes into account fluctuations that do not change the number of crystalline atoms. The argument $t$ emphasizes that all three variables $Q$, $q$, and $f$ evolve in time along the trajectory ${\bf R}(t)$ of the system. 

In an interface pinning simulation, the order parameter $Q$ is computed explicitly, while the variable $q$, which contains the relevant information about the growth of the crystalline region and the motion of the interface, is not directly accessible. To extract information about the interface dynamics from the simulation, we therefore construct a simplified two-dimensional model that separates the effect of $q$ and $f$ on the dynamics of $Q$. In this model, $q$ and $f$ are the sole dynamical variables. We postulate that their dynamics is governed by the effective Hamiltonian 
\begin{equation}
\label{eq:hamilton_effective}
{\mathcal H}(q, f, \dot f)=\alpha q + \frac{\kappa}{2}(f+q-\bar Q)^2 + \frac{\kappa_f}{2} f^2 +\frac{m_f}{2}\dot f^2.
\end{equation}
The first term on the right hand side, $\alpha q$, arises from the imbalance in chemical potential between solid and liquid, which drives the growth (or decrease) of the crystalline region. This term drives the interface motion in the absence of the pinning potential ($\kappa=0$). The effective force $\alpha$ is proportional to the difference in chemical potential, $\mu_{sl}$. The proportionality constant depends on the choice of crystallinity order parameter, Eq. (\ref{eq:alpha}). The second term, $(\kappa/2)(f+q-\bar Q)^2$, couples to $Q=q+f$ and holds the interface fixed near a given position. The parameters $\kappa$ and $\bar Q$ appearing in the pinning potential denote the force constant and the position of the bias, respectively. The third term, $(\kappa_f/2)f^2$, reflects the harmonic potential energy experienced by phonon fluctuations. Finally, the last term is the kinetic energy associated to the time derivative $\dot f$ of the phonon degree of freedom $f$, which carries the effective mass $m_f$. A kinetic energy term for the coordinate $q$ is not necessary, because the dynamics of $q$ is assumed to be overdamped.

\subsection{Langevin equations of motion} 
Based on the effective Hamiltonian of Eq. (\ref{eq:hamilton_effective}) we next devise stochastic equations of motion to describe the coupled time evolution of $q$ and $f$. Since the growth of the interface separating the liquid from the crystalline phase is slow on the molecular time scale, inertial effects in the motion of the variable $q$ are negligible, suggesting to describe its dynamics with an overdamped equation. In contrast, lattice vibrations are fast which requires an underdamped description for the dynamics of $f$. Thus on a coarse-grained level, we expect the dynamics of $q$ and $f$ to be governed by the following pair of coupled Langevin equations,
\begin{align}
\label{eq:Langevin_q}
\gamma \dot q &=-\alpha -\kappa (f + q-\bar Q) + \eta_q(t), \\
m_f \ddot f &= -\kappa_f f - \kappa (f+q-\bar Q)-\gamma_f \dot f+ \eta_f(t).
\label{eq:Langevin_f}
\end{align}
Here, $\gamma$ and $\gamma_f$ are friction constants associated with $q$ and $f$, respectively, and $\eta_q(t)$ and $\eta_f(t)$ are $\delta$-correlated Gaussian random forces. Both the friction and random forces reflects the degrees of freedom neglected in the simplified model \cite{chandlerGreenBook}. The variance of the random forces and the friction constants are related by the fluctuation-dissipation theorem \cite{kubo1966}, $\langle \eta_q(0)\eta_q(t) \rangle = 2 \gamma k_{\rm B}T \delta(t)$ and $\langle \eta_f(0)\eta_f(t) \rangle = 2 \gamma_f k_{\rm B}T \delta(t)$,
% 
%\begin{align}
%\langle \eta_q(0)\eta_q(t) \rangle &= 2 \gamma k_{\rm B}T \delta(t), \\
%\langle \eta_f(0)\eta_f(t) \rangle &= 2 \gamma_f k_{\rm B}T \delta(t), 
%\end{align}
%
where $k_{\rm B}$ is the Boltzmann constant, $T$ is the temperature and the angular brackets $\langle \cdots \rangle$ indicate a thermal average. Solving the above Langevin equations yields $q(t)$ and $f(t)$, from which the time evolution of $Q(t)=q(t)+f(t)$ can be computed.

\subsection{Solution of the stochastic model}
To apply the model to simulation data we first consider the autocorrelation function $\langle \Delta Q(0)\Delta Q(t)\rangle$ where $\Delta Q(t)=Q(t)-\langle Q \rangle$. As shown in details in the Appendix, this autocorrelation function is a sum of three complex exponentials with rather complicated arguments. To obtain a practical expression we consider the long-time limit of the model and utilize the ``separation of time scales''-assumption $\gamma(1/\kappa+1/\kappa_f)\gg\gamma_f/\kappa_f$:
\begin{equation}\label{longtimeQQ}
  \lim_{t\rightarrow\infty}\langle \Delta Q(0)\Delta Q(t) \rangle = \frac{k_BT}{\kappa} A \exp\left(-\frac{t}{\tau}\right)
\end{equation}
where $A= \kappa_f/(\kappa+\kappa_f)$ is the strength
and $\tau = \gamma(\kappa^{-1}+\kappa_f^{-1})$ is the characteristic time of the terminal relaxation. Thus, the friction constant of crystal growth can be determined from a fit to the terminal relaxation time as
\begin{equation}\label{gamma_terminal}
 \gamma = A\tau\kappa.
\end{equation}
The kinetic coefficient of crystal growth $M$ in Eq. (\ref{kc}) is inversely proportional to the friction coefficient $\gamma$. 
The proportionality constant depends on the specifics of the simulations cell used. It is found by first taking the average time-derivative of the definition of $x_s$ schematically depictured in Fig. \ref{ip_scetch}, using that $
 \langle\dot{N}_s\rangle
   =  \langle\dot{q}\rangle\partial \langle N_s\rangle/\partial \langle q\rangle
   =  \langle \dot{q}\rangle N/ Q_{sl}
$ and $\gamma\langle\dot{q}\rangle=-\alpha=-\mu_{sl} N / Q_{sl}$:
% $\langle \dot{x}_s \rangle =- \frac{N^2}{2XY\rho_sQ_{sl}^2\gamma}\mu_{sl}.$
% Thus 
%
\begin{equation} \label{velmu}
M=N^2/2XY\rho_sQ_{sl}^2\gamma.
\end{equation} 
Since $\mu_{sl}$ is known from Eq. (\ref{eq:musl}), the averaged crystal growth rate $\langle \dot x_s \rangle$ can also be deduced.

It is easier to write up the complete solution in the frequency domain. Let us consider the complex admittance, i.e., the frequency dependent mobility $\mu(\omega)$ of $Q$.
%This can be computed from an interface pinned simulation from $Q(t)$ fluctuations using the fluctuation-dissipation theorem (deviation is given in the supplementary material):
%
From the fluctuations of $Q(t)$ it is possible to compute $\mu(\omega)$ using the fluctuation-dissipation theorem \cite{kubo1966,chandlerGreenBook} (see the Appendix):
% Specifically it can be done by a Fourier-Laplace transform of thermal fluctuation of $\dot{Q}$ :
%
\begin{equation}\label{FD}
 \mu(\omega) = \cfrac{\i\omega}{k_BT}\int_0^\infty \langle \Delta Q(0)\Delta \dot{Q}(t)\rangle\exp(-\i\omega t)dt.
\end{equation} 
The complex admittance of the model is (see the Appendix)
\begin{equation}\label{model}
 \mu(\omega) = \left[\left(\cfrac1\gamma+\cfrac{1}{\i \omega m_f+\gamma_f-\i\kappa_f/\omega}\right)^{-1}-\i\kappa/\omega\right]^{-1}.
\end{equation}
In the following subsection we derive this using an analogy to an electric circuit that is mathematically identical to the stochastic model.

\subsection{Electric network representation}

\begin{figure} 
\begin{center} 
  \includegraphics[width=0.7\columnwidth]{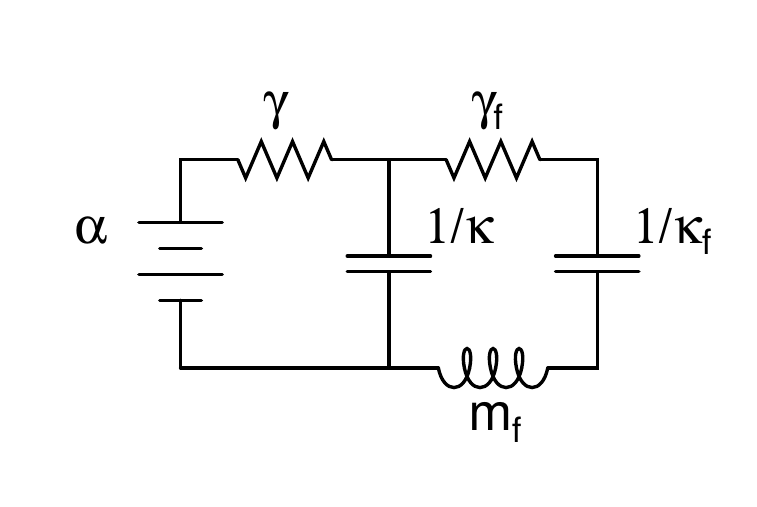} 
  \caption{\label{fig_model} Electric network model corresponding to the effective Hamiltonian and the Langevin equations for the interface motion, Eqs. \ref{eq:hamilton_effective}, \ref{eq:Langevin_q} and \ref{eq:Langevin_f}.} 
\end{center} 
\end{figure}

A system of coupled linear equations of motion can be represented as an idealized electrical circuit. The circuit shown Fig.\ \ref{fig_model} is equivalent to our stochastic model and we can use the rules of electrical networks to derive the solution.
Order parameter values of $Q$, $q'=q-\bar{Q}$ and $f$ are represented by electric charges while voltage drops correspond to forces acting on these order parameters.
The electric circuit consists of capacitors with the admittances $\i\omega/\kappa$ and $\i\omega/\kappa_f$, two resistors with the admittances $1/\gamma$ and $1/\gamma_f$, one inductor with the admittance $-\i/\omega m_f$ and one battery with a voltage $\alpha$. Additionally, the resistors include thermal Johnson-Nyquist noise $\eta_q(t)$ and $\eta_f(t)$. The Langevin equations of motion, Eqs. (\ref{eq:Langevin_q}) and (\ref{eq:Langevin_f}), are retrieved by applying Kirchhoff's voltage law to the left and the right loop, respectively.
The effective Hamiltonian Eq. (\ref{eq:hamilton_effective}) corresponds to the electric energy of the circuit.

The frequency-dependent admittance (Eq.\ (\ref{model})) over the central capacitor can be obtained as follows: The admittances of the unconnected elements from left to right are $\mu_q(\omega)=1/\gamma$, $\mu_Q(\omega)=\i\omega/\kappa$ and $\mu_f(\omega)=1/(\i\omega m_f+\gamma_f-\i\kappa_f/\omega)$. For the latter, we used Ohm's law: impedances (i.e., inverse admittances) connected in series are summed. When currents are running in parallel, however, the admittances that are summed. With respect to the current $\dot Q$ though the central capacitor, the loops on the left and on the right are connected in parallel with each other and in series with the central capacitor. The frequency dependent admittance we are interested in is thus 
\begin{equation}
\mu(\omega)=\left[\frac{1}{\mu_q(\omega)+\mu_f(\omega)}+\frac{1}{\mu_Q(\omega)}\right]^{-1}. 
\end{equation}
By inserting the admittances we arrive at Eq.\ (\ref{model}).

\section{The Lennard-Jones system} \label{secLJ}

To validate the use of the interface pinning method to compute crystal growth rates, we investigate a Lennard-Jones system \cite{lennard-jones1924} of $N=5120$ particles (8$\times$8$\times$20 face centered cubic unit cells; solid-liquid interface in the 100 plane; see Ref.\ \cite{pedersen_ip2013} for details). The potential part of the Hamiltonian is $\sum_{i>j}^Nu(|r_i-r_j|)$ with $u(r)=4\varepsilon((\sigma/r)^{12}-(\sigma/r)^{6})-4\varepsilon((1/2.5)^{12}-(1/2.5)^{6})$ for $r<2.5\sigma$ and zero otherwise. Trajectories are generated with the LAMMPS software package \cite{lammps} using the Verlet integrator with a time step of $0.004\sigma\sqrt{m/\varepsilon}$ where $m$ is the particle mass. The $Np_zT$-ensemble is realized using the Nose-Hoover thermostat \cite{nose1984,hoover1985} with a coupling time of $0.4\sigma\sqrt{m/\varepsilon}$ and the Parrinello-Rahman barostat \cite{parrinello81} with a coupling time of $0.8\sigma\sqrt{m/\varepsilon}$.

%\subsection{Interface pinning}

\begin{figure} 
\begin{center} 
  \includegraphics[width=0.9\columnwidth]{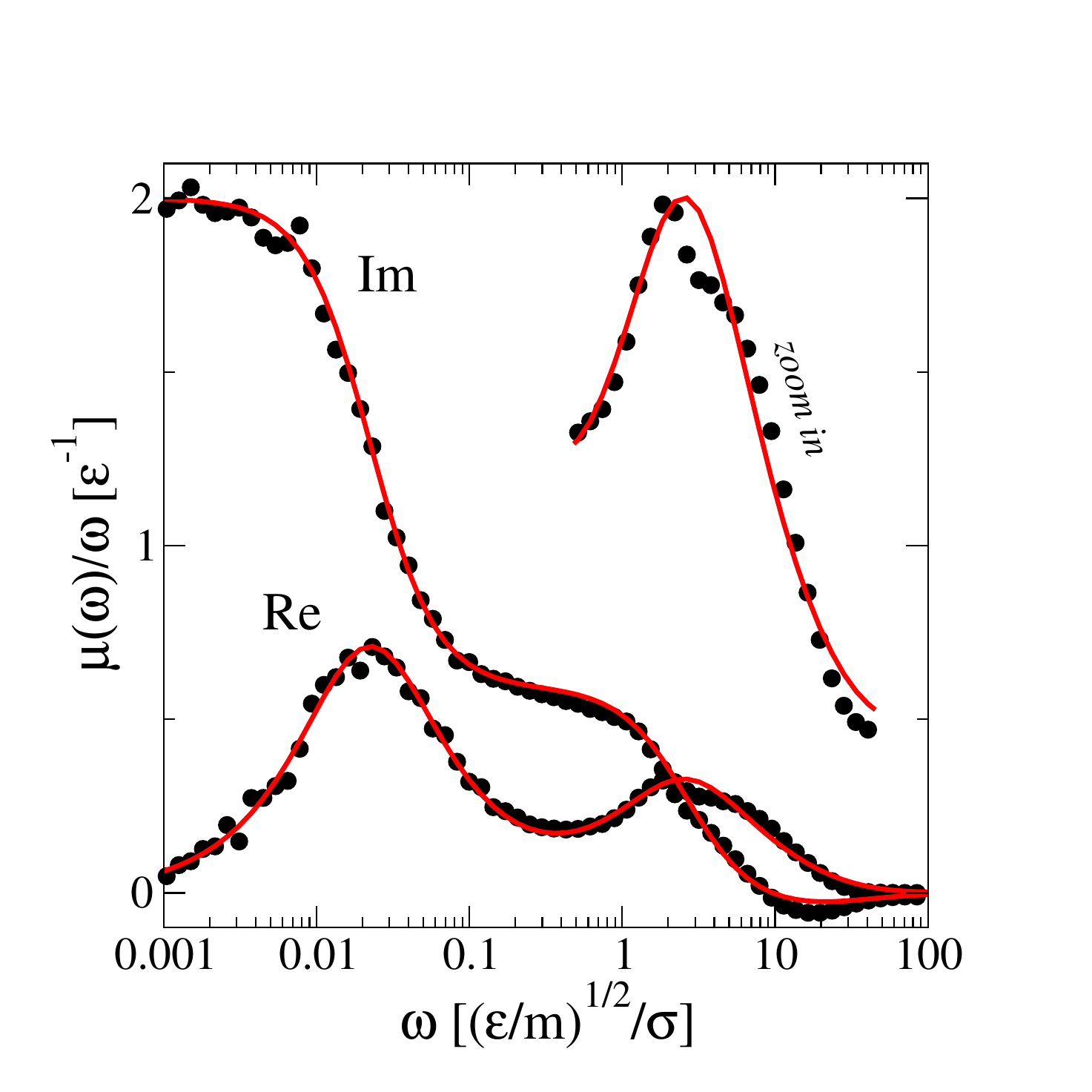} 
\caption{\label{kappa1} Real and imaginary part of $\mu(\omega)/\omega$ in a interface pinning simulation with $\kappa=0.5\varepsilon$. The solid lines correspond to the model shown on Fig. \ref{fig_model} with $\gamma=15.8\sigma\sqrt{m\varepsilon}$, $\kappa_f=1.22\varepsilon$, $\gamma_f=0.70\sigma\sqrt{m\varepsilon}$ and $m_f=0.019m$. The parameters are determined by a least square fit to the real part of $\mu(\omega)/\omega$. The inset zooms in on the high frequencies peak of the real part.} 
\end{center} 
\end{figure}

\begin{figure} 
\begin{center} 
  \includegraphics[width=0.9\columnwidth]{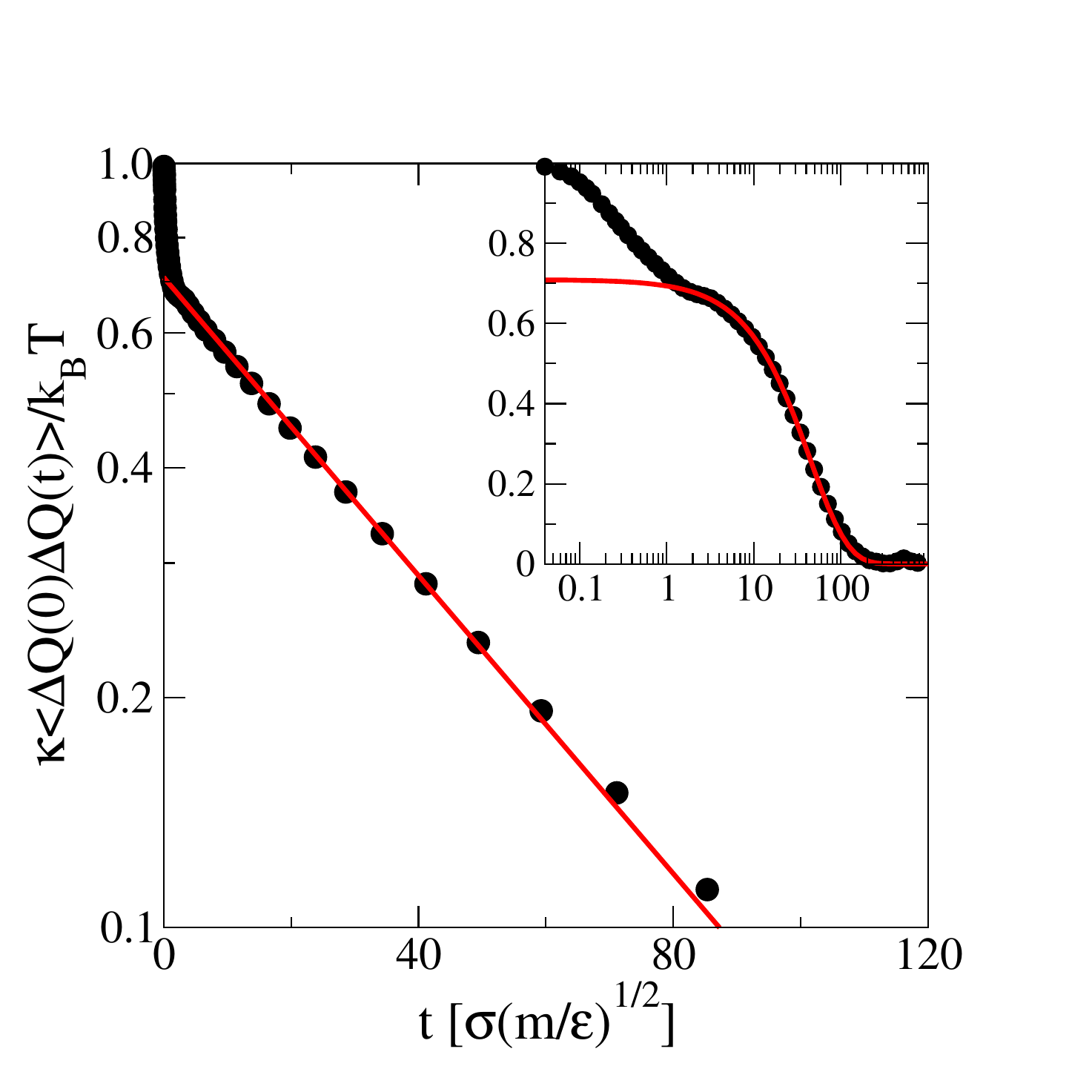}
  \caption{\label{exp} Autocorrelation $\langle \Delta Q(0)\Delta Q(t)\rangle$ from an interface pinning simulation with $\kappa=0.5\varepsilon$ at $T=0.8\varepsilon/k_B$ and the coexistence pressure. The solid red line is the predicted terminal relaxation, Eq. (\ref{longtimeQQ}), using the parameters determined by fitting to $\mu(\omega)/\omega$ (see Fig. \ref{kappa1}).}
\end{center} 
\end{figure}

\begin{figure} 
\begin{center} 
  \includegraphics[width=0.9\columnwidth]{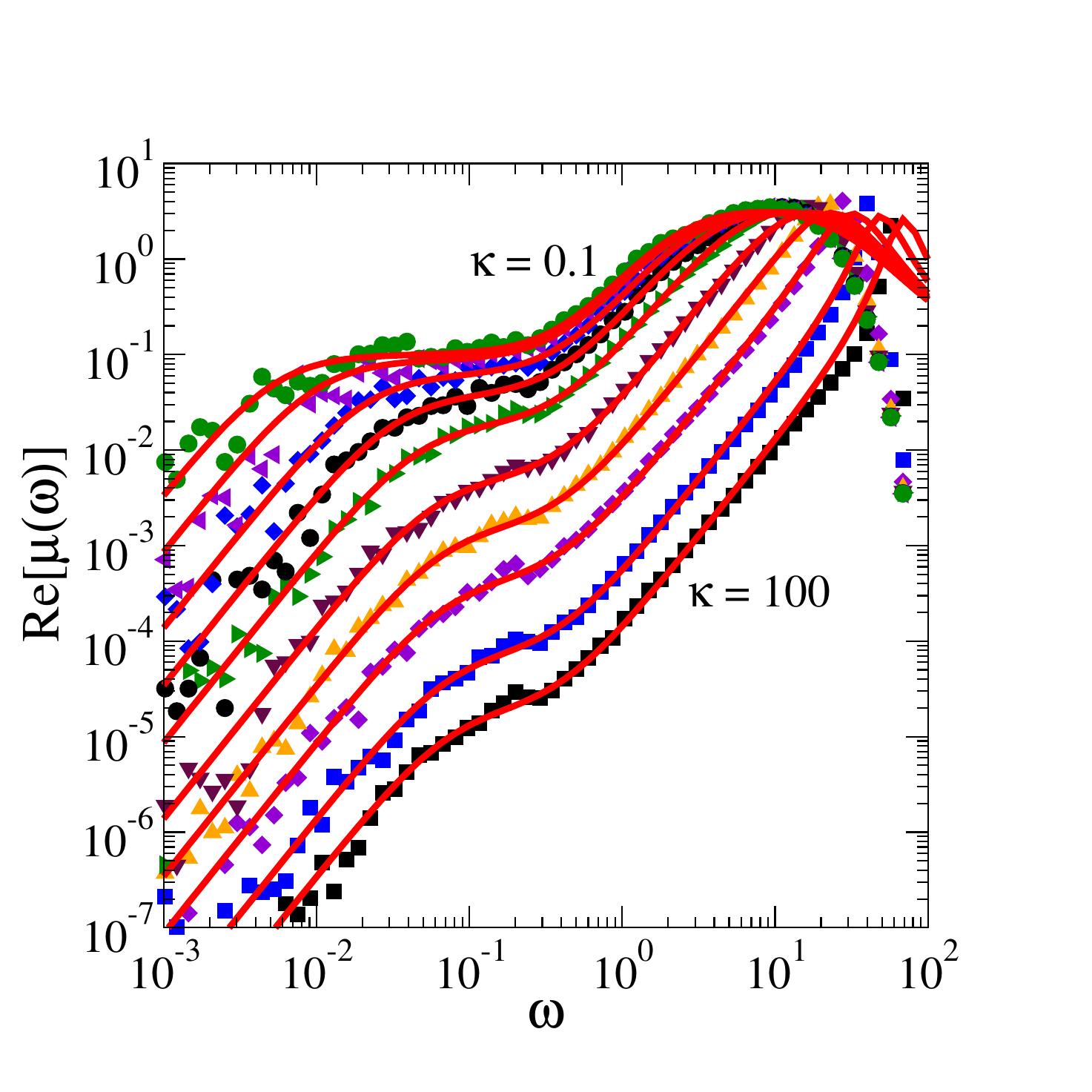} 
  \caption{\label{muall} Real part of $\mu(\omega)$ with $\kappa=\{0.1,0.2,0.5,1,2,5,10,20,50,100\}$ computed using Eq. (\ref{FD}). Solid lines correspond to the model with parameters ($\gamma$, $\gamma_f$, $\kappa_f$ and $m_f$) determined in Fig. \ref{kappa1}.}
\end{center} 
\end{figure}

First we compute $\mu(\omega)$ using Eq. (\ref{FD}). The integral is evaluated numerically from a discrete time series $Q_i$ representing $Q(t)$. The rate $\dot{Q}_i$ is computed from central difference of $Q_i$: $\dot{Q}_i=(Q_{i+1}-Q_{i-1})/(t_{i+1}-t_{i-1})$. Since the autocorrelation function $\langle \Delta Q(0) \Delta \dot{Q}(t)\rangle$ is zero at $t=0$ and $t\rightarrow\infty$ we can replace the Fourier-Laplace transform with a regular Fourier transform, and use the efficient Fast Fourier-Transform (FFT) algorithm to evaluate the integral numerically (we note that the FFT algorithm assumes a periodic dataset and we would get an erroneous result if the integrand did not have the property of vanishing values in the limits). For the analysis, we ensure that the discrete $Q_i$ trajectory has a high sampling frequency so that aliasing is avoided ($t_{i+1}-t_{i}=0.04\sigma\sqrt{m/\varepsilon}$), and are sufficiently long so that slow interface fluctuations are represented.
The dots in Fig. \ref{kappa1} show the real and imaginary part of the computed $\mu(\omega)$ at $T=0.8\varepsilon/k_B$ and the coexistence pressure $p_m=2.185\varepsilon/\sigma^3$ \cite{pedersen_ip2013}. The solid lines shows $\mu(\omega)/\omega$ of our stochastic model with the four parameters $\gamma$, $\gamma_f$, $\kappa_f$ and $m_f$ determined by a least square fit to the real part. The agreement with both the real and imaginary part, which was not used for the fit of the parameters, is excellent. The fit gives $\gamma=15.8\sigma\sqrt{m\varepsilon}$. Fig. \ref{exp} validates that the terminal relaxation time of the autocorrelation function $\langle\Delta Q(0)\Delta Q(t)\rangle$ agrees with this $\gamma$, see Eq. (\ref{longtimeQQ}). Using Eq. (\ref{velmu}) with $N=5120$, $X=Y=12.82\sigma$, $\rho_s=0.973\sigma^\frac{1}{3}$ and $Q_{sl}=56.16$ we arrive at a kinetic coefficient of $M=1.6\sqrt{1/\varepsilon m}$. We note that the model does not give a perfect fit in the high-frequency part. This is, however, not important for the estimate of $\gamma$ as discussed in Sec. \ref{secDiscussion}. Fig. \ref{muall} shows that changing $\kappa$ over three decades yields consistent results with the model parameters determined from the data shown in Fig. \ref{kappa1}.

%\subsection{Free solidification}

\begin{figure} 
\begin{center} 
  \includegraphics[width=0.9\columnwidth]{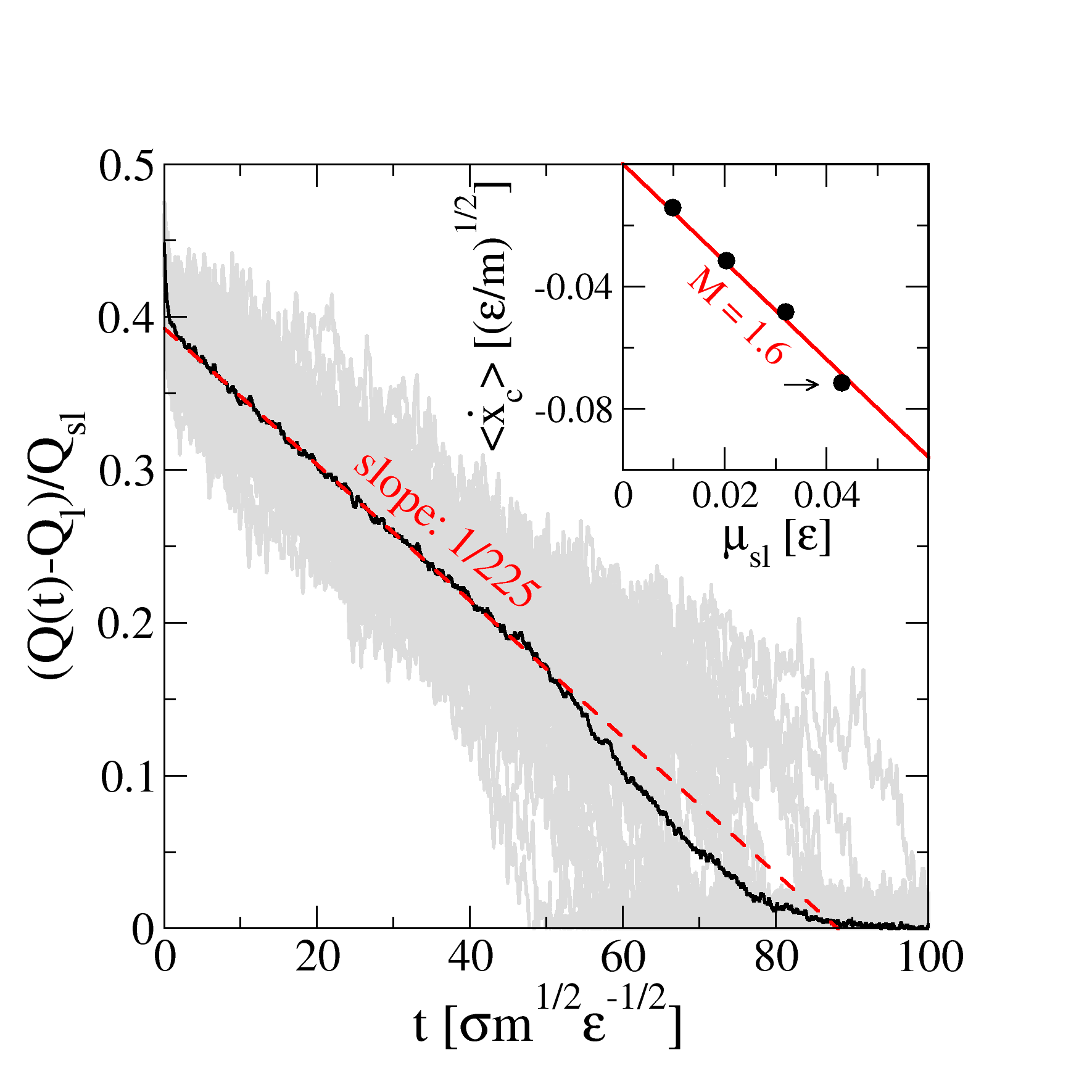} 
 \caption{\label{noUmb} Time evolution of $Q$ along 50 statistically independent LJ melting runs (gray) at $T=0.8\varepsilon/k_B$ and $p=1.8\varepsilon/k_B$ where the initial configurations are taken from equilibrated interface pinning simulations ($\kappa=4\varepsilon$; $\bar{Q}=26$). At this state-point the liquid is the thermodynamically stable phase ($\mu_{sl}=0.043\varepsilon$) and the crystal melts. From the average rate of $\frac{Q(t)-Q_l}{Q_{sl}}$ (red dashed) we compute the average growth velocity to $\langle \dot{x}_s \rangle=-0.0716\sqrt{\varepsilon/m}$ corresponding to $M=1.6\sqrt{1/\varepsilon m}$ (using that $Z_s=32.1\sigma$). The inset shows $\langle \dot{x}_s \rangle$ along the $T=0.8\varepsilon/k_B$ isotherm.}
\end{center} 
\end{figure} 
% Zs = 5120*1.04077/12.869484^2 
%    = 32.1
% ts = 1/0.0044515 
%    = 225
% xs = Zs/ts/2
%    = 0.0716
% mu = 0.0430622
% M  = xs/mu
%    = 1.66

For comparison, we compute $M$ in a direct simulation of crystal melting ($\kappa=0$). The average interface velocity can be computed as
%\begin{equation} \label{DIRECTGamma}
$
\langle \dot{x}_s\rangle=Z_s/2t_s
$
%\end{equation}
where $t_s$ is the average time to crystallize one box length $Z_s=N/\rho_sXY$ (negative values indicate melting). In a simulation, this time can be computed from the average rate of change of the order-parameter
%\begin{equation}
$
1/t_s = \langle \dot{Q} \rangle/Q_{sl}
$.
%\end{equation}
The the solid (black) line in Fig. \ref{noUmb} is the average trajectory when melting a crystal at $T=0.8\varepsilon/k_B$ and $p=1.8\varepsilon/\sigma^3$ (obtained from 50 statistically independent runs). From this result we find that $\langle \dot{x}_s\rangle=-0.0716\sqrt{\varepsilon/m}$. At this state point $\mu_{sl}=0.0431\varepsilon$ and, thus, we get a kinetic coefficient of $M=1.6 \sqrt{1/\varepsilon m}$. This is the same as the value determined by the interface pinning method.

\section{First principle computations} \label{secDFT}

We used the interface pinning method in combination with \textit{ab initio}
simulations to compute the crystal growth rate for real materials. 
In Ref. \cite{physrevB2013} we implemented the interface pinning method
in the Vienna Ab initio Simulation Package (VASP) \cite{kresse1996} and conducted density
functional theory (DFT) calculations of the melting temperature of the period
three elements sodium (Na), magnesium (Mg), aluminum (Al) and silicon (Si).
Without any additional computations, data obtained from these simulations allow us to compute
the kinetic coefficients of these elements in the $z$-direction of the chosen
crystal directions.

For computing the trajectories we employed a Verlet integrator with a timestep
of 4 and 3 fs for Na and the other elements, respectively.
The $Np_zT$-ensemble was realized using a Langevin thermostat  with a coupling
time of 1 ps and a Parinello-Rahman barostat \cite{parrinello81} with a coupling time
of 0.33 ps. PBE and LDA functionals \cite{bloechl1994} were used for the DFT
calculations.

\begin{figure} 
\begin{center} 
  \includegraphics[width=0.9\columnwidth]{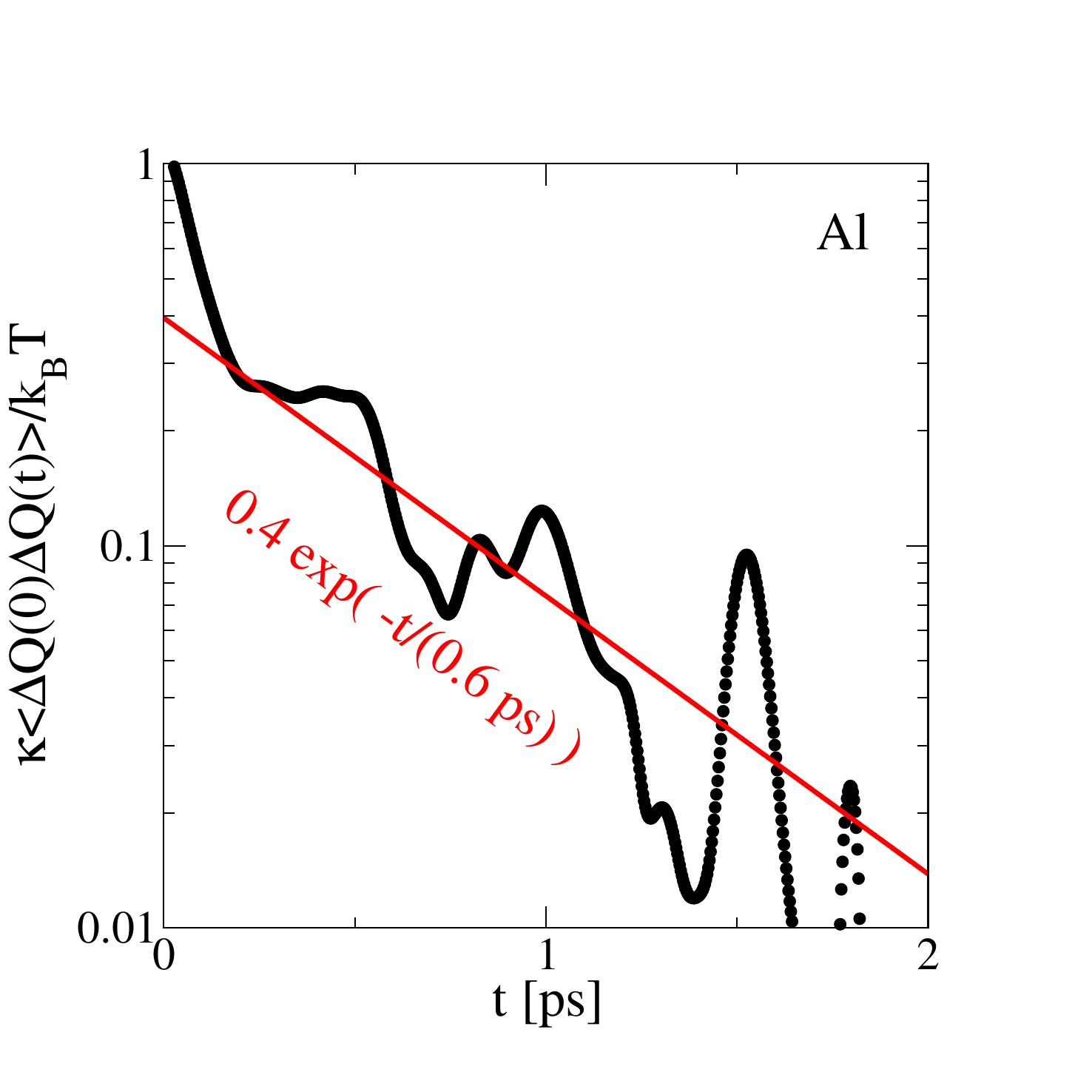} 
  \caption{\label{Al} The $\langle \Delta Q(0)\Delta Q(t)\rangle$ autocorrelation from an {\it ab initio} interface pinning simulation of Al (see Table \ref{tbl_results} and Ref. \cite{physrevB2013} for details). A fit to the terminal exponential relaxation (solid) yields a kinetic coefficient of $M=120$ \AA/(ps eV).} 
\end{center} 
\end{figure}

We compute the kinetic coefficients from the terminal relaxation times of the
autocorrelation function $\langle \Delta Q(0)\Delta Q(t)\rangle$ of the order parameter used for the interface pinning
calculation.
Fig. \ref{Al} shows the autocorrelation function obtained in a first principles
calculation of Al. From a fit to the terminal relaxation we compute $M=120$ \AA/(ps eV) for the
kinetic coefficient in the (100) crystal direction.
Table \ref{tbl_results} lists the computed kinetic coefficients for the
remaining elements along with the direction of crystal growth and the settings
used for the interface pinning calculation.

\begin{table*} 
\begin{tabular}{ c c c | c c c c c c c c c | c }
 element & direction$^\ast$ & unit cells$^\dagger$ &
Q$^\ddagger$ & $\rho_s^{-1}$ [\AA$^3$] & $T_m$ [K] & $N$ & $XY$ [\AA$^2$] &
$Q_{sl}$ & $\kappa$
[eV] & A
& $\tau$ [ps] &
 $M$ [\AA/(ps eV)] \\
\hline

Na     & bcc(100) & 5$\times$5$\times$10 & $Q_6$ & 39.86  & 400  & 500  &
463 & 0.312  & 500 & 0.6  & 1.6
             & 250 \\

Mg     & hcp$(2\bar1\bar10)$ & 4$\times$6$\times$8 & $Q_6$ & 24.60  & 960  & 762  &
583 & 0.244  & 2000 & 0.3  & 2.0
            & 170 \\

Al     & fcc(100) & 4$\times$4$\times$8 & $Q_6$        & 17.59  & 1000 & 512  &
273 & 0.376 & 2000 & 0.4
            & 0.6 &  120 \\

Si     & cd(100) & 3$\times$3$\times$6 & $|\rho_{\bf k}|$   & 20.71  & 1400 & 432  &
271 & 13.9  & 2 & 0.05
            & 2.7 &   130\\

\end{tabular}
\caption{\label{tbl_results}{First principle kinetic coefficients of Na, Mg, Al and Si.}
$^\ast$The direction of crystal growth is given by their Bravais-Miller
indices in the respective crystal system.
Note that we always align the direction of crystal growth with the $z$ axis
in our interface pinning calculation.
$^\dagger$The given numbers refer to the number of orthorhombic unit
cells in each direction in the simulation.
$^\ddagger$We used the rotational
order-parameter $Q_6(\mathbf{R})$ suggested by Steinhardt \cite{steinhardt1983}
and $|\rho_{\mathbf{k}}|$ of Equ. (\ref{eq:rhok}) as order parameters for
the interface pinning calculations}
\end{table*}

\section{Discussion}\label{secDiscussion}

\subsection{Dependence on stochastic model}
We have devised a stochastic model of $Q(t)$ fluctuations in an interface pinning simulation and used it to compute the crystal growth velocity. 
The critical reader might rightfully ask the question: How dependent is the method on the validity of the stochastic model?
To answer this, we emphasize that it is the long-time $\omega\rightarrow 0$ limit of the model that is of importance. If we use a more complicated model of the fast $f(t)$ fluctuations, the long-time solution is the same, except that $\kappa_f$ is replaced by an effective spring constant. To motivate the use of a more complicated model, we note that the high-$\omega$ part of the $\mu(\omega)$ spectrum (inset on Fig. \ref{kincoef}) indicates a superposition of two peaks. The reason for this is that fluctuations in the bulk of the liquid and the crystal are different. Thus, a better model of $Q(t)$ fluctuations would be to split the fast contribution into contributions of the two phases, $f(t)=f_l(t)+f_s(t)$, and write the effective equations of motion as 
\begin{align}
\label{eq:Langevin2_q}
\gamma \dot q &=-\alpha -\kappa (f_l+ f_s + q-\bar Q) + \eta_q(t), \\
m_l \ddot f_l &= - \kappa_l f_l - \kappa (f_l+f_s+q-\bar Q)-\gamma_l \dot f_l+ \eta_{f_l}(t)\\
m_s \ddot f_s &= - \kappa_s f_s - \kappa (f_l+f_s+q-\bar Q)-\gamma_s \dot f_s+ \eta_{f_s}(t).
\label{eq:Langevin2_f}
\end{align}
The frequency dependent mobility (admittance) of this model is
\begin{equation}\label{modelComplicated}
 \mu(\omega) = \left[\frac{1}{\gamma^{-1}+\frac{1}{i \omega m_l+\gamma_l-i\kappa_l/\omega}+\frac{1}{i \omega m_s+\gamma_s-i\kappa_s/\omega}}-i\kappa/\omega\right]^{-1}.
\end{equation}
The long-time limit of this model, however, is the same as in the simpler model we used above, except that $\kappa_f$ is replaced by an effective spring constant $[1/\kappa_l+1/\kappa_s]^{-1}$.

\subsection{Other techniques}
How does the interface pinning method compare to other methods for computing $M$?
To answer this we note that suggested techniques can be grouped into three classes \cite{hoyt2002kc}: (i) free solidification simulations (see Fig. \ref{noUmb}) \cite{huitema1999,laird1992,morris2002}, (ii) fluctuations analysis \cite{briels1997,tepper2002,monk2010} and (iii) forced velocity simulations \cite{broughton1982,sun2004}. This paper's method is of class (ii). The interface pinning method can be viewed as a generalization of the approach suggested by Briels and Tepper \cite{briels1997,tepper2002}. In the Briels-Tepper method two-phase configurations are stabilized by simulating the constant $NVT$ ensemble. This can be viewed as a special case of the interface pinning method, where the order-parameter is the volume, and the bias potential has an infinitely large spring constant. Fluctuations in the number of crystalline particles can be monitored by pressure fluctuations or an order-parameter of crystallinity (e.g. Eq. (\ref{eq:rhok})).
An advantage that the interface pinning method inherits from the Briels-Tepper method is that it involves well-defined equilibrium computations that can be done {\it ad infinitum}.
The following challenges of the Briels-Tepper method are solved with the interface pinning method: 
a) In the Briels-Tepper method, the two-phase configurations are stabilized by keeping the volume constant. In effect, computations can only be done near coexistence. In the interface pinning method the stabilization of two-phase configurations are done by connecting the system with a harmonic field that couples to an order-parameter $Q$ that distinguishes between the crystal and the liquid, and simulations can be performed far into the super cooled or super heated regimes. 
Also, the size of the fluctuations in the number of crystalline particles is determined by the compressibility of the solid and liquid. In effect, they will be large for large systems, leading to long correlation times. With the interface pinning method the size of the $N_s$ fluctuations can be controlled by the value of $\kappa$ and by choosing a good order-parameter making $\kappa_f$ large.
b) $\mu_{sl}$ is directly computed with the interface pinning method. This is an essential property that otherwise had to be computed in separate calculations.

\subsection{Latent heat and volume} 
In this paper we consider the hydrostatic $Np_zT$-ensemble \cite{physrevB2013,pedersen_ip2013} near coexistence. Growth or melting of a solid results in latent heat and latent volume that must be removed to stay in the $Np_zT$-ensemble. 
To avoid that the growth rate is trivially determined by the characteristic time of the thermostat and barostat we investigated crystal growth dynamics with a strong coupling of the thermo- and barostat, i.e., short coupling times. We note that in an weak coupling or experimental situation latent heat and volume disperse away from the interface. In effect, the temperature and pressure at the interface is different from that at the boundaries of the system \cite{sun2004,monk2010}, and in general the temperature and strain tensor fields should be taken explicitly into consideration. Addressing this, goes beyond the scope of this paper.

\subsection{An alternative to forward flux sampling}
We have demonstrated that it is possible to compute the rate of a non-equilibrium process from an equilibrium simulation where an auxiliary harmonic potential is added the Hamiltonian. Auxiliary harmonic potentials are routinely used to compute free energies along reaction coordinates of rare events. Specifically, the {\it umbrella sampling} methods \cite{torrie1977,frenkel2002} uses a series of auxiliary potentials refereed to as {\it umbrellas}. By reweighting \cite{shirts2008} probability distributions generated by the umbrellas it is possible to compute free energies along order parameters. As we have demonstrated here, it is possible to extract also the rates without further computations. We investigated the growth velocity of a flat interface, but presumably the same approach can be used for studying other reaction paths. As an example, by setting up a system with a crystallite \cite{duijneveldt1992,jungblut2013,dittmar2014}, it should be possible to compute the growth velocity of a curved solid-liquid interface. We leave such studies for future investigations.

%\subsection{}
A generalized version of the method may be used for studying dynamics along reaction paths of other rare events, including those of biological systems \cite{duan1998,pedersen2007_hexanol,pianaa2012}. Thus, our method provides an alternative to the forward flux sampling methods \cite{allen2009}. The original version of these methods \cite{allen2005} use a series of interfaces along the path of interest. The rate along a given path is then computed by generating short trajectories between the interfaces. Forward flux sampling is limited to stochastic process; this restriction is not needed with the harmonic pinning approach presented in this paper.

\subsection{Scaling of the kinetic coefficient along the Lennard-Jones melting line}
\begin{figure} 
\begin{center} 
  \includegraphics[width=0.9\columnwidth]{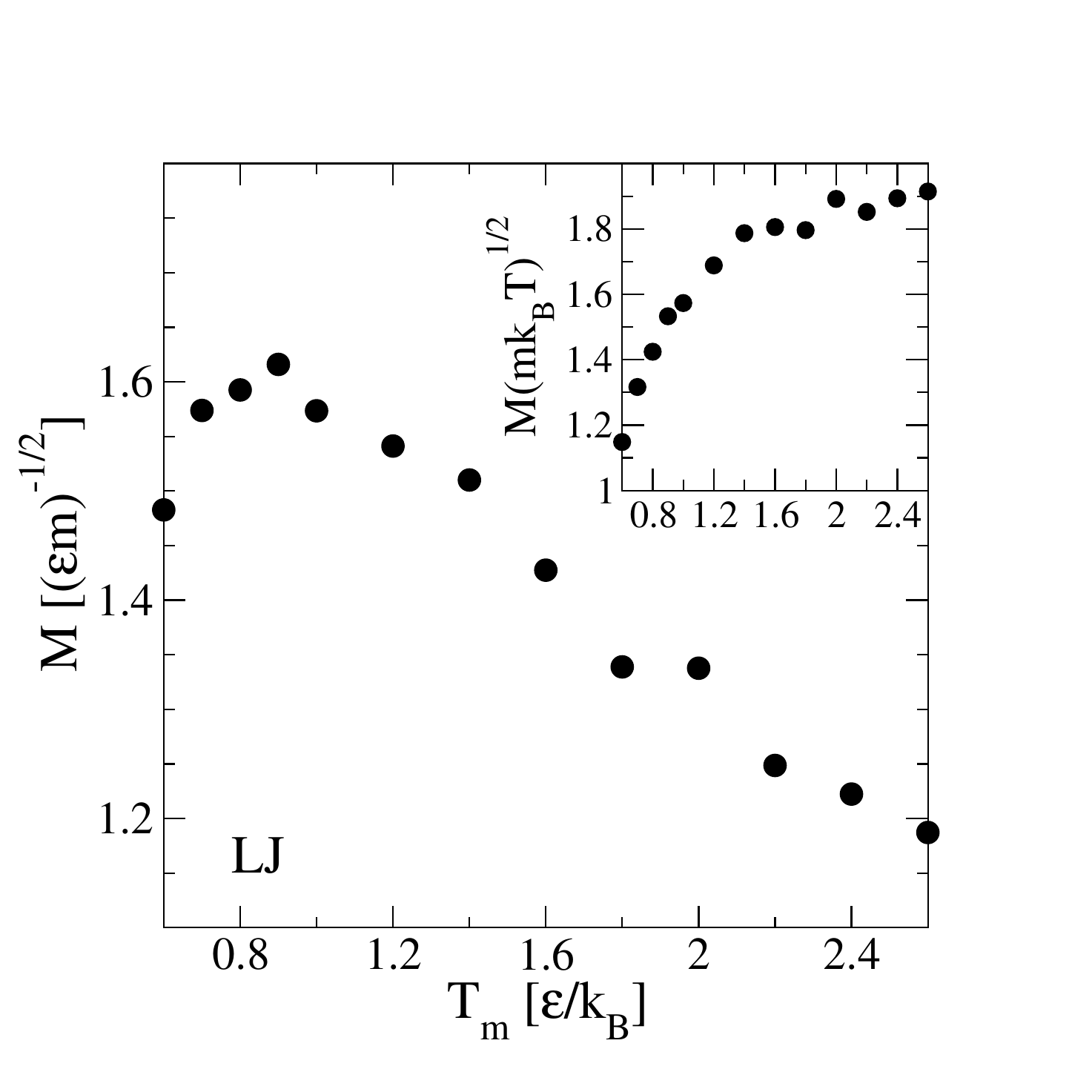} 
  \caption{\label{kincoef} The kinetic coefficient $M$ along the LJ melting line computed with the interface pinning method.} 
\end{center} 
\end{figure} 
Fig. \ref{kincoef} shows the kinetic coefficient $M$ along the coexistence line. In the region $T<1.2$, $M$ is nearly constant. 
At higher temperatures $M$ decreases indicating that the crystal growth (or melting) rate is lower.
The inset show that $M$ in reduced units, using the thermal energy for the energy scale, is roughly invariant at $T>1.2$. In other words, $M$ scales as $1/\sqrt{T}$ along the high temperature (and pressure) part of the melting line. This scale invariance is predicted by the ``isomorph theory'' \cite{gnan2009} of simple liquids \cite{ingebrigtsen2012a}. This theory states that there is a class of system \cite{pedersen_prl2008,scl_I,scl_II}, including the  LJ system \cite{pedersen_prl2008}, that have ``isomorph''-lines in the dense and/or high temperature part of the phase diagram (i.e., not near the critical point). Along these lines structure, dynamics and some thermodynamic properties are invariant in reduced units. One prediction is that the melting line is an isomorph \cite{gnan2009}, and indeed this was shown for the LJ system in Refs. \cite{schroder2011,pedersen_ip2013}. Thus we expect $M$ to be invariant in reduced units along the melting line. This explains why $M$ scales as $1/\sqrt{T}$. The scaling law does not apply near the triple point (note that the lowest temperature data-point on Fig. \ref{kincoef} is at negative pressure). Consistent with this, the melting line itself deviates from an isomorph near the triple point \cite{pedersen_ip2013}. This deviation is probably due to long-ranged attractive interactions \cite{pedersen_ip2013}. We will leave further investigation of isomorph-scale invariance of the crystallization to future studies.

\begin{acknowledgements}
This work was financially supported by the Austrian Science Fund FWF within the SFB ViCoM (F41).
URP was in part funded by the Villum Foundation's grand VKR-023455.
Simulations were carried out on the Vienna Scientific Cluster (VSC).
The authors are grateful for valuable comments and suggestions from Georg Kresse, Jeppe C. Dyre, Lorenzo Costigliola, Tage Christensen and Peter Harrowell.
\end{acknowledgements}

\appendix

\section{Analysis of stochastic model}
In this Appendix we provide a detailed analysis of the stochastic model suggested in the main part of the paper.

\subsection{Static averages and variances} 

The static averages and fluctuations of $q$ and $f$ (and hence also of $Q$) can be computed as averages over the equilibrium distribution 
\begin{equation}
\rho(q, f, \dot f)=\frac{1}{\mathcal Z} \exp\left\{-\beta {\mathcal H}(q, f, \dot f) \right\},
\end{equation}
where $\beta = 1/k_{\rm B}T$ and the partition function 
\begin{equation}
{\mathcal Z}=\int {\rm d}q\, {\rm d}f \,{\rm d}\dot f \exp\left\{-\beta {\mathcal H}(q, f, \dot f) \right\}
\end{equation}
normalizes the distribution. The moments of this multivariate Gaussian distribution can be determined analytically yielding the averages
\begin{align}
\langle q \rangle &=\bar Q -\alpha \left(\frac{1}{\kappa} +\frac{1}{\kappa_f}\right), \\
\langle f \rangle &= \frac{\alpha}{\kappa_f}, \\
\langle \dot f \rangle &= 0,
\end{align}
and variances

\begin{align}
%\langle (\delta q)^2 \rangle &=  \frac{k+\kappa}{\beta k \kappa}\\
\langle (\Delta q)^2 \rangle &=  \frac{1}{\beta}\left(\frac{1}{\kappa}+\frac{1}{\kappa_f}\right)\\
\langle (\Delta f)^2\rangle &=  \langle f^2 \rangle -\langle f \rangle^2= \frac{1}{\beta \kappa_f}, \\
\langle (\Delta \dot f)^2 \rangle &= \langle \dot f^2 \rangle -\langle \dot f \rangle^2= \frac{1}{\beta m_f}.
\end{align}
Here, $\Delta q=q-\langle q\rangle$, $\Delta f=f-\langle f\rangle$ and $\Delta \dot f = \dot f-\langle \dot f\rangle$ are the deviations of $q$, $f$, and $\dot f$ from their respective averages. While $\dot f$ is uncorrelated to the other variables, $q$ and $f$ are correlated with covariance
\begin{equation}
\langle \Delta q \Delta f\rangle =  \langle qf \rangle -\langle q\rangle \langle f \rangle=-\frac{1}{\beta \kappa_f}.
\end{equation} 
From the above expressions, the average and variance of $Q=q+f$ follow,
\begin{align}
\langle Q \rangle &=\bar Q -\frac{\alpha}{\kappa},\\
\langle (\Delta Q)^2\rangle &= \langle Q^2 \rangle -\langle Q \rangle^2=\frac{1}{\beta \kappa}.
\end{align}
Thus, the static fluctuations of $Q$ depend only on the temperature and the force constant $\kappa$ of the pinning potential. 

\subsection{Time correlation functions}

To quantify the average dynamics of the model we now introduce the time correlation functions 
\begin{align}
\phi_{qq}(t) &= \langle \Delta q(0) \Delta q(t) \rangle,\\
\phi_{qf}(t) &= \langle \Delta q(0) \Delta f(t) \rangle, \\
\phi_{f\!f}(t) &= \langle \Delta f(0) \Delta f(t) \rangle.
\end{align}
which correlate the state of the system at time $0$ to its state at a time $t$ later.
It follows from the microscopic reversibility of the equations of motion that $\phi_{qf}(t)=\phi_{fq}(t)$, such that we don't need to consider $\phi_{qf}(t)$ and $\phi_{fq}(t)$ separately.
Thus the time auto correlation functions of the observable order parameter $\langle \Delta Q(0)\Delta Q(t) \rangle$ is
\begin{equation}
\phi_{QQ}(t) = \phi_{qq}(t) + \phi_{f\!f}(t) + 2\phi_{qf}(t).
\end{equation}
One can show, that the following equations hold for the time correlation function $\langle \alpha (0) \beta (t) \rangle$, where $\alpha$ and $\beta$ are two arbitrary functions of $q$ and $f$:
\begin{align}
\frac{d}{dt}\langle \alpha (0) \beta (t) \rangle &=\langle \alpha (0) \dot \beta (t) \rangle=-\langle \dot \alpha (0) \beta (t) \rangle,\\
\frac{d^2}{dt^2}\langle \alpha (0) \beta (t) \rangle&=\langle \alpha (0) \ddot \beta (t) \rangle \nonumber \\
&=-\langle \dot \alpha (0) \dot \beta (t) \rangle=\langle \ddot \alpha (0) \beta (t) \rangle \label{eq:ddot}.
\end{align}
Upon time reversal the time correlation function transform as
\begin{equation}
\langle \alpha (0) \beta (t) \rangle = \varepsilon_\alpha \varepsilon_\beta \langle \alpha (0) \beta (-t) \rangle=\varepsilon_\alpha \varepsilon_\beta \langle \beta (0) \alpha (t) \rangle,
\end{equation}
where $\varepsilon_\alpha$ and $\varepsilon_\beta$ take values $+1$ or $-1$, depending on whether $\alpha$ and $\beta$, respectively, are even or odd under reversal of the momenta. 

By averaging over the equations of motion the following differential equations for the time correlations can be derived,
\begin{align}
\gamma \dot \phi_{qq} &= -\kappa (\phi_{qq}+\phi_{qf}) , \label{eq:phi_qq}\\
\gamma \dot \phi_{qf} &= -\kappa (\phi_{qf}+\phi_{f\!f}), \label{eq:phi_qf}\\
m_f\ddot \phi_{f\!f} &= -(\kappa+\kappa_f) \phi_{f\!f} - \kappa \phi_{qf}-\gamma_f \dot \phi_{f\!f} \label{eq:phi_ff},
\end{align}
where we have omitted the argument of the time correlation functions for simplicity. Introducing the auxiliary function $\psi_{f\!f}(t)=\dot \phi_{f\!f}(t)$, we write these differential equations in matrix notation as
\begin{equation}
\label{eq:ode}
\left(
\begin{array}{c}
\dot \phi_{qq}\\
\dot \phi_{qf} \\
\dot \phi_{f\!f}\\
\dot \psi_{f\!f}
\end{array}
\right)=
%-------------
\left(
\begin{array}{cccc}
-\frac{\kappa}{\gamma} & -\frac{\kappa}{\gamma} & 0 & 0 \\
0 & -\frac{\kappa}{\gamma} & -\frac{\kappa}{\gamma} & 0  \\
0 & 0 & 0 & 1  \\
0 & -\frac{\kappa}{m_f} & -\frac{\kappa+\kappa_f}{m_f} &-\frac{\gamma_f}{m_f} 
\end{array}
\right)
%-------------
\left(
\begin{array}{c}
\phi_{qq}\\
\phi_{qf} \\
\phi_{f\!f}\\
\psi_{f\!f}
\end{array}
\right).
\end{equation}
In a more compact form, this system of homogeneous linear first order differential equations with constant coefficients is expressed as
\begin{equation}
\dot x = A x,
\end{equation}
where $x=\left\{\phi_{qq}, \phi_{qf}, \phi_{f\!f}, \psi_{f\!f}\right\}$ and $A$ is the constant $4\times 4$ matrix on the right hand side of Equ. (\ref{eq:ode}). The formal solution of this equation is given by 
\begin{equation}
x(t) = e^{At}\, x(0)
\end{equation}
with initial conditions 
\begin{equation}
%\label{eq:ode}
x(0)=\left(
\begin{array}{c}
\phi_{qq}(0)\\
\phi_{qf}(0) \\
\phi_{f\!f}(0)\\
\psi_{f\!f}(0)
\end{array}
\right)=\left(
\begin{array}{c}
\langle (\Delta q)^2\rangle\\
\langle \Delta q \Delta f\rangle \\
\langle (\Delta f)^2\rangle\\
\langle \Delta f \Delta \dot f\rangle
\end{array}
\right)=\frac{1}{\beta}
\left(
\begin{array}{c}
 \frac{1}{\kappa}+\frac{1}{\kappa_f}\\
-\frac{1}{\kappa_f} \\
 \frac{1}{\kappa_f}\\
0
\end{array}
\right).
\end{equation}
As can be seen in Equ. (\ref{eq:ode}), the time evolution equations for $\phi_{qf}(t)$, $\phi_{f\!f}(t)$ and $\psi_{f\!f}(t)$ are independent of $\phi_{qq}(t)$. Hence, one can obtain the time correlations functions $\phi_{qf}(t)$, $\phi_{f\!f}(t)$ and $\psi_{f\!f}(t)$  by solving
\begin{equation}
\label{eq:odeB}
\left(
\begin{array}{c}
\dot \phi_{qf} \\
\dot \phi_{f\!f}\\
\dot \psi_{f\!f}
\end{array}
\right)=
%-------------
\left(
\begin{array}{cccc}
-\frac{\kappa}{\gamma} & -\frac{\kappa}{\gamma} & 0  \\
0 & 0 & 1  \\
-\frac{\kappa}{m_f} & -\frac{\kappa+\kappa_f}{m_f} &-\frac{\gamma_f}{m_f} 
\end{array}
\right)
%-------------
\left(
\begin{array}{c}
\phi_{qf} \\
\phi_{f\!f}\\
\psi_{f\!f}
\end{array}
\right).
\end{equation}
As a consequence the three time correlation functions $\phi_{qf}(t)$, $\phi_{f\!f}(t)$ and $\psi_{f\!f}(t)$ can be written as superpositions of three exponentials,
\begin{equation}
x_i(t)=\sum_{j=1}^3 a_{ij} e^{-\lambda_j t},
\end{equation}
where $x_i(t)$ is component $i$ of the time correlation function vector $x(t)$. (Note that we start our numbering at $0$ such that the $0$-th component of $x(t)$ is the time correlation function $\phi_{qq}(t)$.) The time constants $\lambda_i$ are the eigenvalues of the $3 \times 3$ matrix on the right hand side of Equ. (\ref{eq:odeB}) and the specific values of the constants $a_{ij}$ can be determined by diagonalising this matrix and applying the initial conditions  $\phi_{qf}(0)$, $\phi_{f\!f}(0)$ and $\psi_{f\!f}(0)$. The eigenvalues $\lambda_i$, which are also eigenvalues of the matrix $A$ and can be complex for certain parameter combinations, are the roots of the cubic equation
\begin{equation}
\label{eq:cubic}
\lambda^3+\frac{\gamma \gamma_f + \kappa m_f}{\gamma m_f } \lambda^2 +\frac{\gamma \kappa + \gamma_f \kappa + \gamma \kappa_f}{\gamma m_f }\lambda + \frac{\kappa \kappa_f}{\gamma m_f}  = 0.
\end{equation}
Real roots correspond to exponential decay, while complex roots lead to oscillatory behaviour. In general, the time correlation functions $\phi_{qq}(t), \phi_{qf}(t)$, $\phi_{f\!f}(t)$ and $\psi_{f\!f}(t)$ are superpositions of exponential and oscillatory terms. Whether the roots of this equation are real or complex can be determined by computing the discriminant $D$ of the above polynomial in $\lambda$. This polynomial discriminant, as well as the roots of the equation, can be expressed explicitly in terms of the constants $\kappa$, $\kappa_f$, $\gamma$, $\gamma_f$ and $m_f$, but the expressions are omitted here because they are complicated and do not provide much insight. If $D>0$, one root is real and two are complex conjugate, if $D<0$ all roots are real and different, and if $D=0$, all roots are real and at least two of them are equal.  

Once the time correlation function $\phi_{qf}(t)$ is known, $\phi_{qq}(t)$ can be determined by solving the remaining differential equation,
\begin{equation}
\gamma \dot \phi_{qq}(t) = -\kappa (\phi_{qq}(t)+\phi_{qf}(t)).
\end{equation}
Since $\phi_{qf}(t)$ is already given, this equations is an inhomogeneous linear differential equations with constant coefficients, which can be solved, for instance, by variation of constants yielding
\begin{equation}
\phi_{qq}(t)=a_0 e^{-\lambda_0 t} + \sum_{i=1}^3 a_{1i}\frac{\lambda_0}{\lambda_i-\lambda_0}e^{-\lambda_i t}.
\end{equation}
Here, $\lambda_0=\kappa/\gamma$ and the coefficient $a_0$ depends on the initial conditions. Like $\lambda_1$, $\lambda_2$, and $\lambda_3$, also $\lambda_0$ is an eigenvalue of the matrix $A$. Thus, in general, the time correlation function $\phi_{qq}(t)$ is a sum of four exponentials, two of which can lead to oscillatory behaviour. However, explicit solutions of the differential equations for specific parameter sets indicate that the coefficient $a_0$ is close to zero, such that also $\phi_{qq}(t)$ is, effectively, a sum of three exponentials. 

While it is possible to compute the roots of Equ. (\ref{eq:cubic}) analytically, the resulting expressions are exceedingly complicated. For the root closest to zero, which dominates the behaviour of the correlation functions for long times, a simple but accurate approximation can be derived. Neglecting the cubic and quadratic term, the root $\tilde \lambda$ with the smallest magnitude is given by
\begin{equation}
%\tilde \lambda = -\frac{\gamma (\kappa + \kappa_f) + \kappa\gamma_f}{\kappa \kappa_f}.
1/\tilde \lambda = - \gamma\left(\frac{1}{\kappa}+\frac{1}{\kappa_f}\right)-\frac{\gamma_f}{\kappa_f}
\end{equation}
As a further simplification we will use a ``separation of time-scales'' approximation $\gamma(\kappa^{-1}+\kappa_f^{-1})\gg\gamma_f\kappa_f^{-1}$ to eliminate the last term. Thus, the terminal exponential relaxation time is
\begin{equation}
\tau = \gamma\left(\frac{1}{\kappa}+\frac{1}{\kappa_f}\right)
\end{equation}
justifying Eq. (\ref{longtimeQQ}) in the main part of the paper. We note that the term $\tau_f=\gamma_f/\kappa_f$ that we neglect is the characteristic time of the uncoupled ($\kappa=0$) fast $f$ vibrations. For the typical system studied in this paper, this term is less than 1$\%$ of $\tau$.

\subsection{The fluctuation-dissipation theorem and the solution in the frequency domain}\label{app:FD}

According to the fluctuation-dissipation theorem \cite{kubo1966,chandlerGreenBook}, the response of a system to a weak perturbation can be related to the fluctuation properties of the system in equilibrium. To apply this theorem to the pinned interface, imagine a system with the Hamiltonian ${\mathcal H}(z)$ perturbed by a time dependent external field $K(t)$ that couples linearly to the variable $A(z)$. Here, $z$ denotes a set of variables describing the state of the system, for instance the positions and momenta of all particles in our simulation simulation or the variables $Q$, $q$, $f$ and $\dot f$ of our stochastic model. The time-dependent Hamiltonian of the perturbed system is then given by 
\begin{equation}
{\mathcal H}_{\rm pert}(z, t)={\mathcal H}(z)-A(z) K(t).
\end{equation}
The reaction of the system to the external perturbation can be monitored through the change in the variable $B(z)$, which, like $A$, is also a function of $z$, with respect to its equilibrium average,
\begin{equation}
\Delta B(t)=B(t)-\langle B \rangle,
\end{equation}
where the angular brackets denote an equilibrium average and $B(t)$ is a shorthand notation for $B[z(t)]$. Assuming that the external force $K(t)$ has been acting on the system from the infinite past, i.e., from $t=-\infty$, the average deviation $\langle \Delta B(t)\rangle_{\rm ne}$ at time $t$ can be written in its most general form as
\begin{equation}
\langle \Delta B (t)\rangle_{\rm ne}=\int_{-\infty}^t dt'\, K(t') R_{BA}(t-t'),
\end{equation} 
where the response function $R_{BA}(t)$  denotes the response of the system to an impulsive force, i.e., $K(t) \propto \delta (t)$ applied to the system at time $t=0$. The angular brackets $\langle \cdots \rangle_{\rm ne}$ imply an average over many realisations of the the process. According to the fluctuation dissipation theorem, the response function is related to the fluctuation properties of the system by 
\begin{equation}
R_{BA}(t)=\left\{
\begin{array}{ll}
\beta \langle \Delta \dot A(0) \Delta B(t)\rangle & {\rm for}\,\, t\ge0 \\
0 &  {\rm for}\,\, t < 0
\end{array}\right.
\end{equation} 
where $\Delta A(t)=A(t)-\langle A \rangle$ denotes the deviation of $A(t)=A[z(t)]$ from its equilibrium average. Hence, in the linear regime the response of a system to an external perturbation is related to the correlation of spontaneous fluctuations at different times in the equilibrium system. 

The significance of the fluctuation dissipation theorem becomes particularly clear in the frequency domain. In this case, the linear relation between perturbation an response is expressed as 
\begin{equation}
\langle \Delta {\tilde B} (\omega)\rangle_{\rm ne}=\chi_{BA}(\omega) \tilde K(\omega),
\end{equation}
where 
\begin{equation}
\langle \Delta {\tilde B} (\omega)\rangle_{\rm ne}=\int_{-\infty}^{\infty} dt\;  \langle \Delta B (t)\rangle_{\rm ne} e^{-\i\omega t}
\end{equation}
and 
\begin{equation}
\tilde K(\omega) =\int_{-\infty}^{\infty} dt\;  K(t) e^{-\i\omega t}
\end{equation}
are the Fourier transforms of the response and the force, respectively. In the frequency representation the fluctuation-dissipation theorem then links the complex admittance $\chi_{BA}(\omega)$ to the Fourier-Laplace transform of the time correlation function $\langle \Delta \dot A(0) \Delta B(t)\rangle$,
\begin{equation}
\chi_{BA}(\omega)=\beta\int_{0}^{\infty} dt\; \langle \Delta \dot A(0) \Delta B(t)\rangle e^{-\i\omega t}.
\end{equation}
In the following, we will mainly use this frequency-dependent formulation of the fluctuation-dissipation theorem.

We next apply the fluctuation-theorem to our stochastic model with the Hamiltonian of Eq. (\ref{eq:hamilton_effective}) and evolving according to the Langevin equations (\ref{eq:Langevin_q}) and (\ref{eq:Langevin_f}). For this model, we will compute the frequency dependent complex admittance that describes the reaction of the system to an external perturbation acting on the variable $Q=q+f$. The response of the system to the perturbation is monitored using $\dot Q$, the velocity associated with the variable $Q$. So what we would like to determine is the complex admittance $\mu(\omega)$, that relates the average velocity $\dot Q$ to the strength of the external perturbation,
\begin{equation}
\langle \Delta \tilde {\dot Q}(\omega)\rangle_{\rm ne}=\mu(\omega) \tilde K (\omega).
\end{equation}
By comparing the computed admittance $\mu(\omega)$ with results of computer simulations carried out for the atomistic system, we will then determine the values of the model parameters. In particular, we will determine the parameter $\gamma$ which describes the mobility of the interface driven by the difference in chemical potential between the phases. 

For the perturbation $QK(t)=(q+f)K(t)$ the Langevin equations of motion  (\ref{eq:Langevin_q}) and (\ref{eq:Langevin_f}) turn into
\begin{align}
\label{eq:Langevin_pert_q}
\gamma \dot q +\alpha +\kappa (f + q-\bar Q)&= K(t)+\eta_q(t), \\
m_f \ddot f +\gamma_f \dot f + \kappa_f f + \kappa (f+q-\bar Q)&= K(t)+\eta_f(t).
\label{eq:Langevin_pert_f}
\end{align}
Since these equations are linear, we can compute the response of the system analytically for an arbitrarily strong external force $K(t)$. To do that, we carry out a Fourier transformation on the Langevin equations above and average over many realizations of the stochastic process. For $\omega \neq 0$ we obtain
\begin{align}
(\kappa + \i\omega \gamma) \langle \Delta \tilde q\rangle_{\rm ne} + \kappa  \langle \Delta \tilde f\rangle_{\rm ne} &=  \tilde K(\omega),\\
\kappa \langle \Delta \tilde q\rangle_{\rm ne} +(\kappa_f+\kappa-m_f \omega^2+\i\omega \gamma_f) \langle \Delta \tilde f\rangle_{\rm ne} &=  \tilde K(\omega),
\end{align}
where $\Delta \tilde q(\omega)$ and $\Delta \tilde f(\omega)$ are the Fourier transforms of $\Delta q(t) = q(t)-\langle q \rangle$ and $\Delta f(t)= f(t)-\langle f \rangle$, respectively. Note that in the above equations we have omitted the argument $\omega$ in the averages to simplify the notation. Since we are interested in the response of the system in terms of the generalized velocity $\dot Q$, we rewrite these equations for the Fourier transforms of the time derivatives of $\Delta q$ and $\Delta f$,
\begin{align}
\label{eq:Fq}
\frac{\kappa + \i\omega \gamma}{\i\omega} \langle \Delta \tilde {\dot q}\rangle_{\rm ne} + \frac{\kappa}{\i\omega}  \langle \Delta \tilde {\dot f}\rangle_{\rm ne} &=  \tilde K(\omega),\\
\frac{\kappa}{\i\omega}  \langle \Delta \tilde {\dot q}\rangle_{\rm ne} +\frac{\kappa+\kappa_f-m_f \omega^2+\i\omega \gamma_f}{\i\omega} \langle \Delta \tilde {\dot f}\rangle_{\rm ne} &=  \tilde K(\omega),
\label{eq:Ff}
\end{align}
Here we have exploited that in frequency space taking a time derivative simply amounts to multiplication with $\i\omega$, such that $\langle \Delta \tilde {\dot q}\rangle_{\rm ne}=\i\omega \langle \Delta \tilde q\rangle_{\rm ne}$ and $\langle \Delta \tilde {\dot f}\rangle_{\rm ne}=\i\omega \langle \Delta \tilde f\rangle_{\rm ne}$. 

To simplify the notation in the following, we introduce the complex admittances $\mu_q(\omega)$ and $\mu_f(\omega)$ for $\Delta \dot q$ and $\Delta \dot f$ separately in the absence of the coupling term $-\kappa (q+f-\bar Q)^2$ in the Hamiltonian of Eq. (\ref{eq:hamilton_effective}), i.e., without pinning potential. In this case, which corresponds to $\kappa=0$, the equations of motion yield
\begin{align}
 \langle \Delta \tilde {\dot q}\rangle_{\rm ne} &=  \frac{1}{\gamma} \tilde K(\omega),\\
 \langle \Delta \tilde {\dot f}\rangle_{\rm ne} &=  \frac{1}{\gamma_f+\i(m_f \omega-\kappa_f/\omega)}\tilde K(\omega),
\end{align}
such that 
\begin{equation}
\mu_q(\omega)=\frac{1}{\gamma} \hspace{0.3cm}{\rm and}\hspace{0.3cm} \mu_f(\omega)=\frac{1}{\gamma_f+\i(m_f \omega-\kappa_f/\omega)}.
\end{equation}
Using $\mu_q(\omega)$ and $\mu_f(\omega)$,  Eqs. (\ref{eq:Fq}) and (\ref{eq:Ff}) for the system with bias can be written as
\begin{align}
\left(\frac{1}{\mu_q(\omega)}-\frac{\i\kappa}{\omega}\right) \langle \Delta \tilde {\dot q}\rangle_{\rm ne} - \frac{\i\kappa}{\omega}  \langle \Delta \tilde {\dot f}\rangle_{\rm ne} &=  \tilde K(\omega),\\
\left(\frac{1}{\mu_f(\omega)}-\frac{\i\kappa}{\omega}\right)\langle \Delta \tilde {\dot f}\rangle_{\rm ne} -\frac{\i\kappa}{\omega}  \langle \Delta \tilde {\dot q}\rangle_{\rm ne} &=  \tilde K(\omega).
\end{align}
By equating the right hand sides of the above equations one obtains
\begin{equation}
\frac{\langle \Delta \tilde {\dot q}\rangle_{\rm ne}}{\mu_q(\omega)}=\frac{\langle \Delta \tilde {\dot f}\rangle_{\rm ne}}{\mu_f(\omega)},
\end{equation}
implying that 
\begin{align}
\langle \Delta \tilde {\dot q}\rangle_{\rm ne} &=  \frac{\i\omega \mu_q(\omega)}{\i\omega+\kappa\left[\mu_q(\omega)+\mu_f(\omega)\right]}\tilde K(\omega),\\
\langle \Delta \tilde {\dot f}\rangle_{\rm ne} &=  \frac{\i\omega \mu_f(\omega)}{\i\omega+\kappa\left[\mu_q(\omega)+\mu_f(\omega)\right]}\tilde K(\omega).
\end{align}
The response of the system in terms of $\dot Q=\dot q +\dot f$ is then given by 
\begin{equation}
\langle \Delta \tilde {\dot Q}\rangle_{\rm ne} = \frac{\i\omega\left[\mu_q(\omega)+\mu_f(\omega)\right]}{\i\omega+\kappa\left[\mu_q(\omega)+\mu_f(\omega)\right]}\tilde K(\omega),
\end{equation}
such that the complex admittance $\mu(\omega)$ is given by
\begin{equation}
\mu(\omega) = \frac{\mu_q(\omega)+\mu_f(\omega)}{1-\i(\kappa/\omega)\left[\mu_q(\omega)+\mu_f(\omega)\right]}.
\end{equation}
equivalent with Eq. (\ref{model}) in the main part of the paper.
The fluctuation-dissipation theorem links the complex admittance to the Fourier-Laplace transform of the equilibrium time correlation function of $\Delta \dot Q$,
\begin{equation}
\mu(\omega) =\beta \int_0^\infty \langle \Delta \dot Q(0) \Delta \dot Q(t) \rangle e^{-\i\omega t}\; dt.
\label{eq:FD_QQ}
\end{equation}
Thus, the complex admittance $\mu(\omega)$ can be obtained by Fourier-Laplace transformation of the time correlation function $\langle \Delta \dot Q(0) \Delta \dot Q(t) \rangle$ and {\it vice versa} with an inverse Fourier-Laplace transformation.

\bibliography{ipgrowth}

\begin{thebibliography}{43}%
\makeatletter
\providecommand \@ifxundefined [1]{%
 \@ifx{#1\undefined}
}%
\providecommand \@ifnum [1]{%
 \ifnum #1\expandafter \@firstoftwo
 \else \expandafter \@secondoftwo
 \fi
}%
\providecommand \@ifx [1]{%
 \ifx #1\expandafter \@firstoftwo
 \else \expandafter \@secondoftwo
 \fi
}%
\providecommand \natexlab [1]{#1}%
\providecommand \enquote  [1]{``#1''}%
\providecommand \bibnamefont  [1]{#1}%
\providecommand \bibfnamefont [1]{#1}%
\providecommand \citenamefont [1]{#1}%
\providecommand \href@noop [0]{\@secondoftwo}%
\providecommand \href [0]{\begingroup \@sanitize@url \@href}%
\providecommand \@href[1]{\@@startlink{#1}\@@href}%
\providecommand \@@href[1]{\endgroup#1\@@endlink}%
\providecommand \@sanitize@url [0]{\catcode `\\12\catcode `\$12\catcode
  `\&12\catcode `\#12\catcode `\^12\catcode `\_12\catcode `\%12\relax}%
\providecommand \@@startlink[1]{}%
\providecommand \@@endlink[0]{}%
\providecommand \url  [0]{\begingroup\@sanitize@url \@url }%
\providecommand \@url [1]{\endgroup\@href {#1}{\urlprefix }}%
\providecommand \urlprefix  [0]{URL }%
\providecommand \Eprint [0]{\href }%
\providecommand \doibase [0]{http://dx.doi.org/}%
\providecommand \selectlanguage [0]{\@gobble}%
\providecommand \bibinfo  [0]{\@secondoftwo}%
\providecommand \bibfield  [0]{\@secondoftwo}%
\providecommand \translation [1]{[#1]}%
\providecommand \BibitemOpen [0]{}%
\providecommand \bibitemStop [0]{}%
\providecommand \bibitemNoStop [0]{.\EOS\space}%
\providecommand \EOS [0]{\spacefactor3000\relax}%
\providecommand \BibitemShut  [1]{\csname bibitem#1\endcsname}%
\let\auto@bib@innerbib\@empty
%</preamble>
\bibitem [{\citenamefont {Pimpinelli}\ and\ \citenamefont
  {Villain}(1998)}]{pimpinelli1998}%
  \BibitemOpen
  \bibfield  {author} {\bibinfo {author} {\bibfnamefont {A.}~\bibnamefont
  {Pimpinelli}}\ and\ \bibinfo {author} {\bibfnamefont {J.}~\bibnamefont
  {Villain}},\ }\href@noop {} {\emph {\bibinfo {title} {Physics of Crystal
  Growth}}}\ (\bibinfo  {publisher} {Cambridge University Press},\ \bibinfo
  {year} {1998})\BibitemShut {NoStop}%
\bibitem [{\citenamefont {Woodruff}(1973)}]{woodruff1973}%
  \BibitemOpen
  \bibfield  {author} {\bibinfo {author} {\bibfnamefont {D.~P.}\ \bibnamefont
  {Woodruff}},\ }\href@noop {} {\emph {\bibinfo {title} {The Solid-Liquid
  Interface}}}\ (\bibinfo  {publisher} {Cambridge University Press},\ \bibinfo
  {year} {1973})\BibitemShut {NoStop}%
\bibitem [{\citenamefont {Provatas}\ and\ \citenamefont
  {Elder}(2010)}]{provatas2010}%
  \BibitemOpen
  \bibfield  {author} {\bibinfo {author} {\bibfnamefont {N.}~\bibnamefont
  {Provatas}}\ and\ \bibinfo {author} {\bibfnamefont {K.}~\bibnamefont
  {Elder}},\ }\href@noop {} {\emph {\bibinfo {title} {Phase-field methods in
  materials science and Engineering}}}\ (\bibinfo  {publisher} {WILEY-VCH},\
  \bibinfo {year} {2010})\BibitemShut {NoStop}%
\bibitem [{\citenamefont {Hoyt}\ \emph {et~al.}(2002)\citenamefont {Hoyt},
  \citenamefont {Asta},\ and\ \citenamefont {Karma}}]{hoyt2002kc}%
  \BibitemOpen
  \bibfield  {author} {\bibinfo {author} {\bibfnamefont {J.~J.}\ \bibnamefont
  {Hoyt}}, \bibinfo {author} {\bibfnamefont {M.}~\bibnamefont {Asta}}, \ and\
  \bibinfo {author} {\bibfnamefont {A.}~\bibnamefont {Karma}},\ }\href@noop {}
  {\bibfield  {journal} {\bibinfo  {journal} {Interface Science}\ }\textbf
  {\bibinfo {volume} {10}},\ \bibinfo {pages} {181} (\bibinfo {year}
  {2002})}\BibitemShut {NoStop}%
\bibitem [{\citenamefont {Hoyt}\ and\ \citenamefont {Asta}(2002)}]{hoyt2002}%
  \BibitemOpen
  \bibfield  {author} {\bibinfo {author} {\bibfnamefont {J.~J.}\ \bibnamefont
  {Hoyt}}\ and\ \bibinfo {author} {\bibfnamefont {M.}~\bibnamefont {Asta}},\
  }\href {\doibase 10.1103/PhysRevB.65.214106} {\bibfield  {journal} {\bibinfo
  {journal} {Phys. Rev. B}\ }\textbf {\bibinfo {volume} {65}},\ \bibinfo
  {pages} {214106} (\bibinfo {year} {2002})}\BibitemShut {NoStop}%
\bibitem [{\citenamefont {Kubo}(1966)}]{kubo1966}%
  \BibitemOpen
  \bibfield  {author} {\bibinfo {author} {\bibfnamefont {R.}~\bibnamefont
  {Kubo}},\ }\href {\doibase 10.1088/0034-4885/29/1/306} {\bibfield  {journal}
  {\bibinfo  {journal} {Rep. Prog. Phys.}\ }\textbf {\bibinfo {volume} {29}},\
  \bibinfo {pages} {255} (\bibinfo {year} {1966})}\BibitemShut {NoStop}%
\bibitem [{\citenamefont {Pedersen}\ \emph {et~al.}(2013)\citenamefont
  {Pedersen}, \citenamefont {Hummel}, \citenamefont {Kresse}, \citenamefont
  {Kahl},\ and\ \citenamefont {Dellago}}]{physrevB2013}%
  \BibitemOpen
  \bibfield  {author} {\bibinfo {author} {\bibfnamefont {U.~R.}\ \bibnamefont
  {Pedersen}}, \bibinfo {author} {\bibfnamefont {F.}~\bibnamefont {Hummel}},
  \bibinfo {author} {\bibfnamefont {G.}~\bibnamefont {Kresse}}, \bibinfo
  {author} {\bibfnamefont {G.}~\bibnamefont {Kahl}}, \ and\ \bibinfo {author}
  {\bibfnamefont {C.}~\bibnamefont {Dellago}},\ }\href {\doibase
  10.1063/1.4818747} {\bibfield  {journal} {\bibinfo  {journal} {Phys. Rev. B}\
  }\textbf {\bibinfo {volume} {88}},\ \bibinfo {pages} {094101} (\bibinfo
  {year} {2013})}\BibitemShut {NoStop}%
\bibitem [{\citenamefont {Pedersen}(2013)}]{pedersen_ip2013}%
  \BibitemOpen
  \bibfield  {author} {\bibinfo {author} {\bibfnamefont {U.~R.}\ \bibnamefont
  {Pedersen}},\ }\href {\doibase 10.1103/PhysRevB.88.094101} {\bibfield
  {journal} {\bibinfo  {journal} {J. Chem Phys.}\ }\textbf {\bibinfo {volume}
  {139}},\ \bibinfo {pages} {104102} (\bibinfo {year} {2013})}\BibitemShut
  {NoStop}%
\bibitem [{\citenamefont {Langevin}(1908)}]{langevin1908}%
  \BibitemOpen
  \bibfield  {author} {\bibinfo {author} {\bibfnamefont {P.}~\bibnamefont
  {Langevin}},\ }\href@noop {} {\bibfield  {journal} {\bibinfo  {journal} {C.
  R. Acad. Sci. (Paris)}\ }\textbf {\bibinfo {volume} {146}},\ \bibinfo {pages}
  {530Ð533} (\bibinfo {year} {1908})}\BibitemShut {NoStop}%
\bibitem [{\citenamefont {Chandler}(1987)}]{chandlerGreenBook}%
  \BibitemOpen
  \bibfield  {author} {\bibinfo {author} {\bibfnamefont {D.}~\bibnamefont
  {Chandler}},\ }\href@noop {} {\emph {\bibinfo {title} {Introduction to Modern
  Statistical Mechanics}}}\ (\bibinfo  {publisher} {Oxford University Press},\
  \bibinfo {year} {1987})\BibitemShut {NoStop}%
\bibitem [{\citenamefont {Lennard-Jones}(1924)}]{lennard-jones1924}%
  \BibitemOpen
  \bibfield  {author} {\bibinfo {author} {\bibfnamefont {J.~E.}\ \bibnamefont
  {Lennard-Jones}},\ }\href@noop {} {\bibfield  {journal} {\bibinfo  {journal}
  {Proc. R. Soc. Lond. A}\ }\textbf {\bibinfo {volume} {106}},\ \bibinfo
  {pages} {463} (\bibinfo {year} {1924})}\BibitemShut {NoStop}%
\bibitem [{\citenamefont {Plimpton}(1995)}]{lammps}%
  \BibitemOpen
  \bibfield  {author} {\bibinfo {author} {\bibfnamefont {S.~J.}\ \bibnamefont
  {Plimpton}},\ }\href {http://lammps.sandia.gov} {\bibfield  {journal}
  {\bibinfo  {journal} {J. Comp. Phys.}\ }\textbf {\bibinfo {volume} {117}},\
  \bibinfo {pages} {1} (\bibinfo {year} {1995})}\BibitemShut {NoStop}%
\bibitem [{\citenamefont {Nose}(1984)}]{nose1984}%
  \BibitemOpen
  \bibfield  {author} {\bibinfo {author} {\bibfnamefont {S.}~\bibnamefont
  {Nose}},\ }\href {\doibase 10.1063/1.447334} {\bibfield  {journal} {\bibinfo
  {journal} {J. Chem. Phys.}\ }\textbf {\bibinfo {volume} {81}},\ \bibinfo
  {pages} {511} (\bibinfo {year} {1984})}\BibitemShut {NoStop}%
\bibitem [{\citenamefont {Hoover}(1985)}]{hoover1985}%
  \BibitemOpen
  \bibfield  {author} {\bibinfo {author} {\bibfnamefont {W.~G.}\ \bibnamefont
  {Hoover}},\ }\href {\doibase 10.1103/PhysRevA.31.1695} {\bibfield  {journal}
  {\bibinfo  {journal} {Phys. Rev. A}\ }\textbf {\bibinfo {volume} {31}},\
  \bibinfo {pages} {1695} (\bibinfo {year} {1985})}\BibitemShut {NoStop}%
\bibitem [{\citenamefont {Parrinello}\ and\ \citenamefont
  {Rahman}(1981)}]{parrinello81}%
  \BibitemOpen
  \bibfield  {author} {\bibinfo {author} {\bibfnamefont {M.}~\bibnamefont
  {Parrinello}}\ and\ \bibinfo {author} {\bibfnamefont {A.}~\bibnamefont
  {Rahman}},\ }\href {\doibase 10.1063/1.328693} {\bibfield  {journal}
  {\bibinfo  {journal} {J. Appl. Phys.}\ }\textbf {\bibinfo {volume} {52}},\
  \bibinfo {pages} {7182} (\bibinfo {year} {1981})}\BibitemShut {NoStop}%
\bibitem [{\citenamefont {Kresse}\ and\ \citenamefont
  {Furthm\"uller}(1996)}]{kresse1996}%
  \BibitemOpen
  \bibfield  {author} {\bibinfo {author} {\bibfnamefont {G.}~\bibnamefont
  {Kresse}}\ and\ \bibinfo {author} {\bibfnamefont {J.}~\bibnamefont
  {Furthm\"uller}},\ }\href {\doibase 10.1103/PhysRevB.54.11169} {\bibfield
  {journal} {\bibinfo  {journal} {Phys. Rev. B}\ }\textbf {\bibinfo {volume}
  {54}},\ \bibinfo {pages} {11169} (\bibinfo {year} {1996})}\BibitemShut
  {NoStop}%
\bibitem [{\citenamefont {Bl\"ochl}(1994)}]{bloechl1994}%
  \BibitemOpen
  \bibfield  {author} {\bibinfo {author} {\bibfnamefont {P.~E.}\ \bibnamefont
  {Bl\"ochl}},\ }\href {\doibase 10.1103/PhysRevB.50.17953} {\bibfield
  {journal} {\bibinfo  {journal} {Phys. Rev. B}\ }\textbf {\bibinfo {volume}
  {50}},\ \bibinfo {pages} {17953} (\bibinfo {year} {1994})}\BibitemShut
  {NoStop}%
\bibitem [{\citenamefont {Steinhardt}\ \emph {et~al.}(1983)\citenamefont
  {Steinhardt}, \citenamefont {Nelson},\ and\ \citenamefont
  {Ronchetti}}]{steinhardt1983}%
  \BibitemOpen
  \bibfield  {author} {\bibinfo {author} {\bibfnamefont {P.~J.}\ \bibnamefont
  {Steinhardt}}, \bibinfo {author} {\bibfnamefont {D.~R.}\ \bibnamefont
  {Nelson}}, \ and\ \bibinfo {author} {\bibfnamefont {M.}~\bibnamefont
  {Ronchetti}},\ }\href {\doibase 10.1103/PhysRevB.28.784} {\bibfield
  {journal} {\bibinfo  {journal} {Phys. Rev. B}\ }\textbf {\bibinfo {volume}
  {28}},\ \bibinfo {pages} {784} (\bibinfo {year} {1983})}\BibitemShut
  {NoStop}%
\bibitem [{\citenamefont {Huitema}\ \emph {et~al.}(1999)\citenamefont
  {Huitema}, \citenamefont {Vlot},\ and\ \citenamefont {{van der
  Eerden}}}]{huitema1999}%
  \BibitemOpen
  \bibfield  {author} {\bibinfo {author} {\bibfnamefont {H.~E.~A.}\
  \bibnamefont {Huitema}}, \bibinfo {author} {\bibfnamefont {M.~J.}\
  \bibnamefont {Vlot}}, \ and\ \bibinfo {author} {\bibfnamefont {J.~P.}\
  \bibnamefont {{van der Eerden}}},\ }\href {\doibase 10.1063/1.479233}
  {\bibfield  {journal} {\bibinfo  {journal} {J. Chem Phys.}\ }\textbf
  {\bibinfo {volume} {111}},\ \bibinfo {pages} {4714} (\bibinfo {year}
  {1999})}\BibitemShut {NoStop}%
\bibitem [{\citenamefont {Laird}\ and\ \citenamefont
  {Haymet}(1992)}]{laird1992}%
  \BibitemOpen
  \bibfield  {author} {\bibinfo {author} {\bibfnamefont {B.~B.}\ \bibnamefont
  {Laird}}\ and\ \bibinfo {author} {\bibfnamefont {A.~D.~J.}\ \bibnamefont
  {Haymet}},\ }\href {\doibase 10.1021/cr00016a007} {\bibfield  {journal}
  {\bibinfo  {journal} {Chem. Rev.}\ }\textbf {\bibinfo {volume} {92}},\
  \bibinfo {pages} {1819} (\bibinfo {year} {1992})}\BibitemShut {NoStop}%
\bibitem [{\citenamefont {Morris}\ and\ \citenamefont
  {Song}(2002)}]{morris2002}%
  \BibitemOpen
  \bibfield  {author} {\bibinfo {author} {\bibfnamefont {J.~R.}\ \bibnamefont
  {Morris}}\ and\ \bibinfo {author} {\bibfnamefont {X.}~\bibnamefont {Song}},\
  }\href {\doibase 10.1063/1.1474581} {\bibfield  {journal} {\bibinfo
  {journal} {J. Chem. Phys.}\ }\textbf {\bibinfo {volume} {116}},\ \bibinfo
  {pages} {9352} (\bibinfo {year} {2002})}\BibitemShut {NoStop}%
\bibitem [{\citenamefont {Briels}\ and\ \citenamefont
  {Tepper}(1997)}]{briels1997}%
  \BibitemOpen
  \bibfield  {author} {\bibinfo {author} {\bibfnamefont {W.~J.}\ \bibnamefont
  {Briels}}\ and\ \bibinfo {author} {\bibfnamefont {H.~L.}\ \bibnamefont
  {Tepper}},\ }\href {\doibase 10.1103/PhysRevLett.79.5074} {\bibfield
  {journal} {\bibinfo  {journal} {Phys. Rev. Lett.}\ }\textbf {\bibinfo
  {volume} {79}},\ \bibinfo {pages} {5074} (\bibinfo {year}
  {1997})}\BibitemShut {NoStop}%
\bibitem [{\citenamefont {Tepper}\ and\ \citenamefont
  {Briels}(2002)}]{tepper2002}%
  \BibitemOpen
  \bibfield  {author} {\bibinfo {author} {\bibfnamefont {H.~L.}\ \bibnamefont
  {Tepper}}\ and\ \bibinfo {author} {\bibfnamefont {W.~J.}\ \bibnamefont
  {Briels}},\ }\href {\doibase 10.1063/1.1452110} {\bibfield  {journal}
  {\bibinfo  {journal} {J. Chem. Phys.}\ }\textbf {\bibinfo {volume} {116}},\
  \bibinfo {pages} {5186} (\bibinfo {year} {2002})}\BibitemShut {NoStop}%
\bibitem [{\citenamefont {Monk}\ \emph {et~al.}(2010)\citenamefont {Monk},
  \citenamefont {Yang}, \citenamefont {Mendelev}, \citenamefont {Asta},
  \citenamefont {Hoyt},\ and\ \citenamefont {Sun}}]{monk2010}%
  \BibitemOpen
  \bibfield  {author} {\bibinfo {author} {\bibfnamefont {J.}~\bibnamefont
  {Monk}}, \bibinfo {author} {\bibfnamefont {Y.}~\bibnamefont {Yang}}, \bibinfo
  {author} {\bibfnamefont {M.~I.}\ \bibnamefont {Mendelev}}, \bibinfo {author}
  {\bibfnamefont {M.}~\bibnamefont {Asta}}, \bibinfo {author} {\bibfnamefont
  {J.~J.}\ \bibnamefont {Hoyt}}, \ and\ \bibinfo {author} {\bibfnamefont
  {D.~Y.}\ \bibnamefont {Sun}},\ }\href {\doibase
  10.1088/0965-0393/18/1/015004} {\bibfield  {journal} {\bibinfo  {journal}
  {Modelling Simul. Mater. Sci. Eng.}\ }\textbf {\bibinfo {volume} {18}},\
  \bibinfo {pages} {015004} (\bibinfo {year} {2010})}\BibitemShut {NoStop}%
\bibitem [{\citenamefont {Broughton}\ \emph {et~al.}(1982)\citenamefont
  {Broughton}, \citenamefont {Gilmer},\ and\ \citenamefont
  {Jackson}}]{broughton1982}%
  \BibitemOpen
  \bibfield  {author} {\bibinfo {author} {\bibfnamefont {J.~Q.}\ \bibnamefont
  {Broughton}}, \bibinfo {author} {\bibfnamefont {G.~H.}\ \bibnamefont
  {Gilmer}}, \ and\ \bibinfo {author} {\bibfnamefont {K.~A.}\ \bibnamefont
  {Jackson}},\ }\href {\doibase 10.1103/PhysRevLett.49.1496} {\bibfield
  {journal} {\bibinfo  {journal} {Phys. Rev. Lett.}\ }\textbf {\bibinfo
  {volume} {49}},\ \bibinfo {pages} {1496} (\bibinfo {year}
  {1982})}\BibitemShut {NoStop}%
\bibitem [{\citenamefont {Sun}\ \emph {et~al.}(2004)\citenamefont {Sun},
  \citenamefont {Asta},\ and\ \citenamefont {Hoyt}}]{sun2004}%
  \BibitemOpen
  \bibfield  {author} {\bibinfo {author} {\bibfnamefont {D.~Y.}\ \bibnamefont
  {Sun}}, \bibinfo {author} {\bibfnamefont {M.}~\bibnamefont {Asta}}, \ and\
  \bibinfo {author} {\bibfnamefont {J.~J.}\ \bibnamefont {Hoyt}},\ }\href
  {\doibase 10.1103/PhysRevB.69.024108} {\bibfield  {journal} {\bibinfo
  {journal} {Phys. Rev. B}\ }\textbf {\bibinfo {volume} {69}},\ \bibinfo
  {pages} {024108} (\bibinfo {year} {2004})}\BibitemShut {NoStop}%
\bibitem [{\citenamefont {Torrie}\ and\ \citenamefont
  {Valleau}(1997)}]{torrie1977}%
  \BibitemOpen
  \bibfield  {author} {\bibinfo {author} {\bibfnamefont {G.}~\bibnamefont
  {Torrie}}\ and\ \bibinfo {author} {\bibfnamefont {J.}~\bibnamefont
  {Valleau}},\ }\href {\doibase 10.1016/0021-9991(77)90121-8} {\bibfield
  {journal} {\bibinfo  {journal} {J. Comp. Phys.}\ }\textbf {\bibinfo {volume}
  {23}},\ \bibinfo {pages} {187} (\bibinfo {year} {1997})}\BibitemShut
  {NoStop}%
\bibitem [{\citenamefont {Frenkel}\ and\ \citenamefont
  {Smit}(2002)}]{frenkel2002}%
  \BibitemOpen
  \bibfield  {author} {\bibinfo {author} {\bibfnamefont {D.}~\bibnamefont
  {Frenkel}}\ and\ \bibinfo {author} {\bibfnamefont {B.}~\bibnamefont {Smit}},\
  }\href@noop {} {\emph {\bibinfo {title} {Understanding Molecular Simulation:
  From Algorithms to Applications}}},\ \bibinfo {edition} {2nd}\ ed.,\ edited
  by\ \bibinfo {editor} {\bibfnamefont {D.}~\bibnamefont {Frenkel}}, \bibinfo
  {editor} {\bibfnamefont {M.}~\bibnamefont {Klein}}, \bibinfo {editor}
  {\bibfnamefont {M.}~\bibnamefont {Parrinello}}, \ and\ \bibinfo {editor}
  {\bibfnamefont {B.}~\bibnamefont {Smit}},\ \bibinfo {series} {Computational
  Science Series}, Vol.~\bibinfo {volume} {1}\ (\bibinfo  {publisher} {Academic
  Press},\ \bibinfo {year} {2002})\BibitemShut {NoStop}%
\bibitem [{\citenamefont {Shirts}\ and\ \citenamefont
  {Chodera}(2008)}]{shirts2008}%
  \BibitemOpen
  \bibfield  {author} {\bibinfo {author} {\bibfnamefont {M.~R.}\ \bibnamefont
  {Shirts}}\ and\ \bibinfo {author} {\bibfnamefont {J.~D.}\ \bibnamefont
  {Chodera}},\ }\href {\doibase 10.1063/1.2978177} {\bibfield  {journal}
  {\bibinfo  {journal} {J. Chem Phys.}\ }\textbf {\bibinfo {volume} {129}},\
  \bibinfo {pages} {124105} (\bibinfo {year} {2008})}\BibitemShut {NoStop}%
\bibitem [{\citenamefont {van Duijneveldt}\ and\ \citenamefont
  {Frenkel}(1992)}]{duijneveldt1992}%
  \BibitemOpen
  \bibfield  {author} {\bibinfo {author} {\bibfnamefont {J.~S.}\ \bibnamefont
  {van Duijneveldt}}\ and\ \bibinfo {author} {\bibfnamefont {D.}~\bibnamefont
  {Frenkel}},\ }\href {\doibase 10.1063/1.462802} {\bibfield  {journal}
  {\bibinfo  {journal} {J. Chem. Phys.}\ }\textbf {\bibinfo {volume} {96}},\
  \bibinfo {pages} {4655} (\bibinfo {year} {1992})}\BibitemShut {NoStop}%
\bibitem [{\citenamefont {Jungblut}\ \emph {et~al.}(2013)\citenamefont
  {Jungblut}, \citenamefont {Singraber},\ and\ \citenamefont
  {Dellago}}]{jungblut2013}%
  \BibitemOpen
  \bibfield  {author} {\bibinfo {author} {\bibfnamefont {S.}~\bibnamefont
  {Jungblut}}, \bibinfo {author} {\bibfnamefont {A.}~\bibnamefont {Singraber}},
  \ and\ \bibinfo {author} {\bibfnamefont {C.}~\bibnamefont {Dellago}},\ }\href
  {\doibase 10.1080/00268976.2013.832820} {\bibfield  {journal} {\bibinfo
  {journal} {Mol. Phys.}\ }\textbf {\bibinfo {volume} {111}},\ \bibinfo {pages}
  {22} (\bibinfo {year} {2013})}\BibitemShut {NoStop}%
\bibitem [{\citenamefont {Dittmar}\ and\ \citenamefont
  {Kusalik}(2014)}]{dittmar2014}%
  \BibitemOpen
  \bibfield  {author} {\bibinfo {author} {\bibfnamefont {H.}~\bibnamefont
  {Dittmar}}\ and\ \bibinfo {author} {\bibfnamefont {P.~G.}\ \bibnamefont
  {Kusalik}},\ }\href {\doibase 10.1103/PhysRevLett.112.195701} {\bibfield
  {journal} {\bibinfo  {journal} {Phys. Rev. Lett.}\ }\textbf {\bibinfo
  {volume} {112}},\ \bibinfo {pages} {195701} (\bibinfo {year}
  {2014})}\BibitemShut {NoStop}%
\bibitem [{\citenamefont {Duan}\ and\ \citenamefont
  {Kollman}(1998)}]{duan1998}%
  \BibitemOpen
  \bibfield  {author} {\bibinfo {author} {\bibfnamefont {Y.}~\bibnamefont
  {Duan}}\ and\ \bibinfo {author} {\bibfnamefont {P.~A.}\ \bibnamefont
  {Kollman}},\ }\href {\doibase 10.1126/science.282.5389.740} {\bibfield
  {journal} {\bibinfo  {journal} {Science}\ }\textbf {\bibinfo {volume}
  {282}},\ \bibinfo {pages} {740} (\bibinfo {year} {1998})}\BibitemShut
  {NoStop}%
\bibitem [{\citenamefont {Pedersen}\ \emph {et~al.}(2007)\citenamefont
  {Pedersen}, \citenamefont {Peters},\ and\ \citenamefont
  {Westh}}]{pedersen2007_hexanol}%
  \BibitemOpen
  \bibfield  {author} {\bibinfo {author} {\bibfnamefont {U.~R.}\ \bibnamefont
  {Pedersen}}, \bibinfo {author} {\bibfnamefont {G.~H.}\ \bibnamefont
  {Peters}}, \ and\ \bibinfo {author} {\bibfnamefont {P.}~\bibnamefont
  {Westh}},\ }\href {\doibase 10.1016/j.bpc.2006.07.005} {\bibfield  {journal}
  {\bibinfo  {journal} {Biophys. Chem.}\ }\textbf {\bibinfo {volume} {125}},\
  \bibinfo {pages} {104} (\bibinfo {year} {2007})}\BibitemShut {NoStop}%
\bibitem [{\citenamefont {Piana}\ \emph {et~al.}(2012)\citenamefont {Piana},
  \citenamefont {Lindorff-Larsen},\ and\ \citenamefont {Shaw}}]{pianaa2012}%
  \BibitemOpen
  \bibfield  {author} {\bibinfo {author} {\bibfnamefont {S.}~\bibnamefont
  {Piana}}, \bibinfo {author} {\bibfnamefont {K.}~\bibnamefont
  {Lindorff-Larsen}}, \ and\ \bibinfo {author} {\bibfnamefont {D.~E.}\
  \bibnamefont {Shaw}},\ }\href {\doibase 10.1073/pnas.1201811109} {\bibfield
  {journal} {\bibinfo  {journal} {PNAS}\ }\textbf {\bibinfo {volume} {109}},\
  \bibinfo {pages} {17845} (\bibinfo {year} {2012})}\BibitemShut {NoStop}%
\bibitem [{\citenamefont {Allen}\ \emph {et~al.}(2009)\citenamefont {Allen},
  \citenamefont {Valeriani},\ and\ \citenamefont {ten Wolde}}]{allen2009}%
  \BibitemOpen
  \bibfield  {author} {\bibinfo {author} {\bibfnamefont {R.~J.}\ \bibnamefont
  {Allen}}, \bibinfo {author} {\bibfnamefont {C.}~\bibnamefont {Valeriani}}, \
  and\ \bibinfo {author} {\bibfnamefont {P.~R.}\ \bibnamefont {ten Wolde}},\
  }\href {\doibase 10.1088/0953-8984/21/46/463102} {\bibfield  {journal}
  {\bibinfo  {journal} {J. Phys. Condens. Matter}\ }\textbf {\bibinfo {volume}
  {21}},\ \bibinfo {pages} {463102} (\bibinfo {year} {2009})}\BibitemShut
  {NoStop}%
\bibitem [{\citenamefont {Allen}\ \emph {et~al.}(2005)\citenamefont {Allen},
  \citenamefont {Warren},\ and\ \citenamefont {ten Wolde}}]{allen2005}%
  \BibitemOpen
  \bibfield  {author} {\bibinfo {author} {\bibfnamefont {R.~J.}\ \bibnamefont
  {Allen}}, \bibinfo {author} {\bibfnamefont {P.~B.}\ \bibnamefont {Warren}}, \
  and\ \bibinfo {author} {\bibfnamefont {P.~R.}\ \bibnamefont {ten Wolde}},\
  }\href@noop {} {\bibfield  {journal} {\bibinfo  {journal} {Phys. Rev. Lett.}\
  }\textbf {\bibinfo {volume} {94}},\ \bibinfo {pages} {018104} (\bibinfo
  {year} {2005})}\BibitemShut {NoStop}%
\bibitem [{\citenamefont {Gnan}\ \emph {et~al.}(2009)\citenamefont {Gnan},
  \citenamefont {Schr{\o}der}, \citenamefont {Pedersen}, \citenamefont
  {Bailey},\ and\ \citenamefont {Dyre}}]{gnan2009}%
  \BibitemOpen
  \bibfield  {author} {\bibinfo {author} {\bibfnamefont {N.}~\bibnamefont
  {Gnan}}, \bibinfo {author} {\bibfnamefont {T.~B.}\ \bibnamefont
  {Schr{\o}der}}, \bibinfo {author} {\bibfnamefont {U.~R.}\ \bibnamefont
  {Pedersen}}, \bibinfo {author} {\bibfnamefont {N.~P.}\ \bibnamefont
  {Bailey}}, \ and\ \bibinfo {author} {\bibfnamefont {J.~C.}\ \bibnamefont
  {Dyre}},\ }\href {\doibase 10.1063/1.3265957} {\bibfield  {journal} {\bibinfo
   {journal} {J. Chem. Phys.}\ }\textbf {\bibinfo {volume} {131}},\ \bibinfo
  {pages} {234504} (\bibinfo {year} {2009})}\BibitemShut {NoStop}%
\bibitem [{\citenamefont {Ingebrigtsen}\ \emph {et~al.}(2012)\citenamefont
  {Ingebrigtsen}, \citenamefont {Schr{\o}der},\ and\ \citenamefont
  {Dyre}}]{ingebrigtsen2012a}%
  \BibitemOpen
  \bibfield  {author} {\bibinfo {author} {\bibfnamefont {T.~S.}\ \bibnamefont
  {Ingebrigtsen}}, \bibinfo {author} {\bibfnamefont {T.~B.}\ \bibnamefont
  {Schr{\o}der}}, \ and\ \bibinfo {author} {\bibfnamefont {J.~C.}\ \bibnamefont
  {Dyre}},\ }\href {\doibase 10.1103/PhysRevX.2.011011} {\bibfield  {journal}
  {\bibinfo  {journal} {Phys. Rev. X}\ }\textbf {\bibinfo {volume} {2}},\
  \bibinfo {pages} {011011} (\bibinfo {year} {2012})}\BibitemShut {NoStop}%
\bibitem [{\citenamefont {Pedersen}\ \emph {et~al.}(2008)\citenamefont
  {Pedersen}, \citenamefont {Bailey}, \citenamefont {Schr{\o}der},\ and\
  \citenamefont {Dyre}}]{pedersen_prl2008}%
  \BibitemOpen
  \bibfield  {author} {\bibinfo {author} {\bibfnamefont {U.~R.}\ \bibnamefont
  {Pedersen}}, \bibinfo {author} {\bibfnamefont {N.~P.}\ \bibnamefont
  {Bailey}}, \bibinfo {author} {\bibfnamefont {T.~B.}\ \bibnamefont
  {Schr{\o}der}}, \ and\ \bibinfo {author} {\bibfnamefont {J.~C.}\ \bibnamefont
  {Dyre}},\ }\href {\doibase 10.1103/PhysRevLett.100.015701} {\bibfield
  {journal} {\bibinfo  {journal} {Phys. Rev. Lett.}\ }\textbf {\bibinfo
  {volume} {100}},\ \bibinfo {pages} {015701} (\bibinfo {year}
  {2008})}\BibitemShut {NoStop}%
\bibitem [{\citenamefont {Bailey}\ \emph
  {et~al.}(2008{\natexlab{a}})\citenamefont {Bailey}, \citenamefont {Pedersen},
  \citenamefont {Gnan}, \citenamefont {Schr¿der},\ and\ \citenamefont
  {Dyre}}]{scl_I}%
  \BibitemOpen
  \bibfield  {author} {\bibinfo {author} {\bibfnamefont {N.~P.}\ \bibnamefont
  {Bailey}}, \bibinfo {author} {\bibfnamefont {U.~R.}\ \bibnamefont
  {Pedersen}}, \bibinfo {author} {\bibfnamefont {N.}~\bibnamefont {Gnan}},
  \bibinfo {author} {\bibfnamefont {T.~B.}\ \bibnamefont {Schr¿der}}, \ and\
  \bibinfo {author} {\bibfnamefont {J.~C.}\ \bibnamefont {Dyre}},\ }\href
  {\doibase 10.1063/1.2982247} {\bibfield  {journal} {\bibinfo  {journal} {J.
  Chem. Phys.}\ }\textbf {\bibinfo {volume} {129}},\ \bibinfo {pages} {184507}
  (\bibinfo {year} {2008}{\natexlab{a}})}\BibitemShut {NoStop}%
\bibitem [{\citenamefont {Bailey}\ \emph
  {et~al.}(2008{\natexlab{b}})\citenamefont {Bailey}, \citenamefont {Pedersen},
  \citenamefont {Gnan}, \citenamefont {Schr¿der},\ and\ \citenamefont
  {Dyre}}]{scl_II}%
  \BibitemOpen
  \bibfield  {author} {\bibinfo {author} {\bibfnamefont {N.~P.}\ \bibnamefont
  {Bailey}}, \bibinfo {author} {\bibfnamefont {U.~R.}\ \bibnamefont
  {Pedersen}}, \bibinfo {author} {\bibfnamefont {N.}~\bibnamefont {Gnan}},
  \bibinfo {author} {\bibfnamefont {T.~B.}\ \bibnamefont {Schr¿der}}, \ and\
  \bibinfo {author} {\bibfnamefont {J.~C.}\ \bibnamefont {Dyre}},\ }\href
  {\doibase 10.1063/1.2982249} {\bibfield  {journal} {\bibinfo  {journal} {J.
  Chem. Phys.}\ }\textbf {\bibinfo {volume} {129}},\ \bibinfo {pages} {184508}
  (\bibinfo {year} {2008}{\natexlab{b}})}\BibitemShut {NoStop}%
\bibitem [{\citenamefont {Schr{\o}der}\ \emph {et~al.}(2011)\citenamefont
  {Schr{\o}der}, \citenamefont {Gnan}, \citenamefont {Pedersen}, \citenamefont
  {Bailey},\ and\ \citenamefont {Dyre}}]{schroder2011}%
  \BibitemOpen
  \bibfield  {author} {\bibinfo {author} {\bibfnamefont {T.~B.}\ \bibnamefont
  {Schr{\o}der}}, \bibinfo {author} {\bibfnamefont {N.}~\bibnamefont {Gnan}},
  \bibinfo {author} {\bibfnamefont {U.~R.}\ \bibnamefont {Pedersen}}, \bibinfo
  {author} {\bibfnamefont {N.~P.}\ \bibnamefont {Bailey}}, \ and\ \bibinfo
  {author} {\bibfnamefont {J.~C.}\ \bibnamefont {Dyre}},\ }\href {\doibase
  10.1063/1.3582900} {\bibfield  {journal} {\bibinfo  {journal} {The Journal of
  Chemical Physics}\ }\textbf {\bibinfo {volume} {134}},\ \bibinfo {eid}
  {164505} (\bibinfo {year} {2011})}\BibitemShut {NoStop}%
\end{thebibliography}%

\end{document}